\documentclass[pre,onecolumn,preprintnumbers,showpacs,floatfix,amsmath,amssymb]{revtex4}
\usepackage{graphicx}
\usepackage{natbib}
\usepackage{subfigure}
\usepackage{color}
\usepackage{multirow}

\graphicspath{{figs/}}

\begin{document}

\title{Drops with non-circular footprints}
\author{Pablo D. Ravazzoli, Alejandro G. Gonz\'alez, Javier A. Diez }
\affiliation{Instituto de F\'{\i}sica Arroyo Seco (CIFICEN-CONICET), Universidad Nacional del Centro de la Provincia de Buenos Aires, Pinto 399, 7000, Tandil, Argentina}

\begin{abstract}
In this paper we study the morphology of drops formed on partially wetting substrates, whose footprint is not circular. This type of drops is a consequence of the breakup processes occurring in thin films when anisotropic contact line motions take place. The anisotropy is basically due to hysteresis effects of the contact angle since some parts of the contact line are wetting, while others are dewetting. Here, we obtain a peculiar drop shape from the rupture of a long liquid filament sitting on a solid substrate, and analyze its shape and contact angles by means of goniometric and refractive techniques. We also find a non--trivial steady state solution for the drop shape within the long wave approximation (lubrication theory), and compare most of its features with experimental data. This solution is presented both in Cartesian and polar coordinates, whose constants must be determined by a certain group of measured parameters. Besides, we obtain the dynamics of the drop generation from numerical simulations of the full Navier--Stokes equation, where we emulate the hysteretic effects with an appropriate spatial distribution of the static contact angle over the substrate.
\end{abstract}

\maketitle

\section{Introduction}
\label{sec:intro}

Wetting of solids, such as the spreading of a drop on a surface, is a basic and ubiquitous phenomenon in a wide variety of natural and technological processes. As a consequence, a large amount of intensive experimental and theoretical research has been published over the two past decades~\cite{deGennesRMP,berg_93,degennes02}. Here, we focus on the shape of the footprint of a drop. In general, it is assumed that the sessile drop on a horizontal substrate has a circular footprint~\cite{michael_prsl77}. However, this not usually the case, as can be easily seen by just spreading some water over the table. The reasons for having non--circular shapes are primarily due to non-uniform wettability of the substrate or/and contact angle hysteresis~\cite{Joanny84,gao_06,eral_13}. The former effect is usually attributed to variations of physico--chemical properties of the surface coating, such as its energy, which can yield to different contact angles. On the other hand, the later stands for the range of static contact angles that a circular sessile drop can attain on a substrate with uniform wettability. However, these effects are usually mixed in the literature into a single one, simply named as contact angle hysteresis~\cite{schwartz_lang98,schwartz98,opik_jcis_00}. 

In many cases, the wetting phenomena are complicated mainly due to this hysteretic behavior of the contact angle. Actually, different angles can be observed along the same contact line at rest even if the surface is smooth. Their minimum and maximum are called receding and advancing contact angles, respectively. Thus, bearing in mind this fact, it is not surprising that the evolution of the liquid spreading from an arbitrary initial condition to its final state at rest can actually play a significant role on the final drop shape~\cite{mahady_15}. Non axisymmetric drops can also be obtained from other mechanisms, such as the coalescence of two sessile drops~\cite{zhang_pof15}.

In this work, we are concerned with the shape of drops that form as a consequence of the breakup of a liquid filament sitting on a horizontal substrate under partial wetting conditions. We study the main features of these drops, such as footprints, contact angle distribution around their peripheries, and thickness profiles along the longitudinal and transverse directions to the filament.

The paper is organized as follows. In Section~\ref{sec:exp0} we present the experimental part by characterizing the substrate and the liquid to be used, as well as the rupture mechanism of the filament and the resulting droplets. In section~\ref{sec:form} we describe the formalism of the long wave approximation which gives the framework for the static solution to be obtained. As a consequence, the description is firstly performed under the assumption of small contact angles. Thus, Sections~\ref{sec:lub_cart} and  \ref{sec:lub_polar} are devoted to the development of analytical solutions for non--axisymmetric sessile drops, i.e. with  non-circular footprints, as well for their comparison with experimental data. Solutions without the restriction of small contact angles are obtained numerically in Section~\ref{sec:num}, where the Navier--Stokes equations are solved to describe the dewetting dynamics of a filament of finite length, which ends up into a single droplet. Finally, in Section~\ref{sec:exp1} we compare the experimental, theoretical, and numerical results for the footprint of the drop, contact angles and thickness profiles. In Section~\ref{sec:conc}, we summarize the results and present the concluding remarks.

\section{Drops from a filament breakup: Experiments}
\label{sec:exp0}

\subsection{Substrate and liquid}

In order to perform the experiments, we use a substrate that is partially wetted by our working fluid, namely a silicon oil (polydimethylsiloxane, PDMS). 
The substrate is a microscope slide (glass) which is coated with a fluorinated solution (EGC-1700 of 3M) by dip coating under controlled speed using a Chemat Dip Coater. This process ensures that the PDMS partially wets the substrate, since the coating lowers the surface energy of the glass. The surface tension, $\gamma$, and density, $\rho$, of PDMS are measured with a Kr\"uss K11 tensiometer, while its viscosity, $\nu $, is measured with a Haake VT550 rotating viscometer. The values of the measured parameters are: $\gamma = 19.8\, dyn/cm$, $\rho = 0.96~g/cm^{3}$, and $\nu = 20~ St$ at temperature $T=20^\circ C$, which yield a capillary length $a_c=\sqrt{\gamma/(\rho \,g)}=1.45~mm$, where $g$ is gravity.

\begin{figure}
\centering
\includegraphics[width=0.8\linewidth]{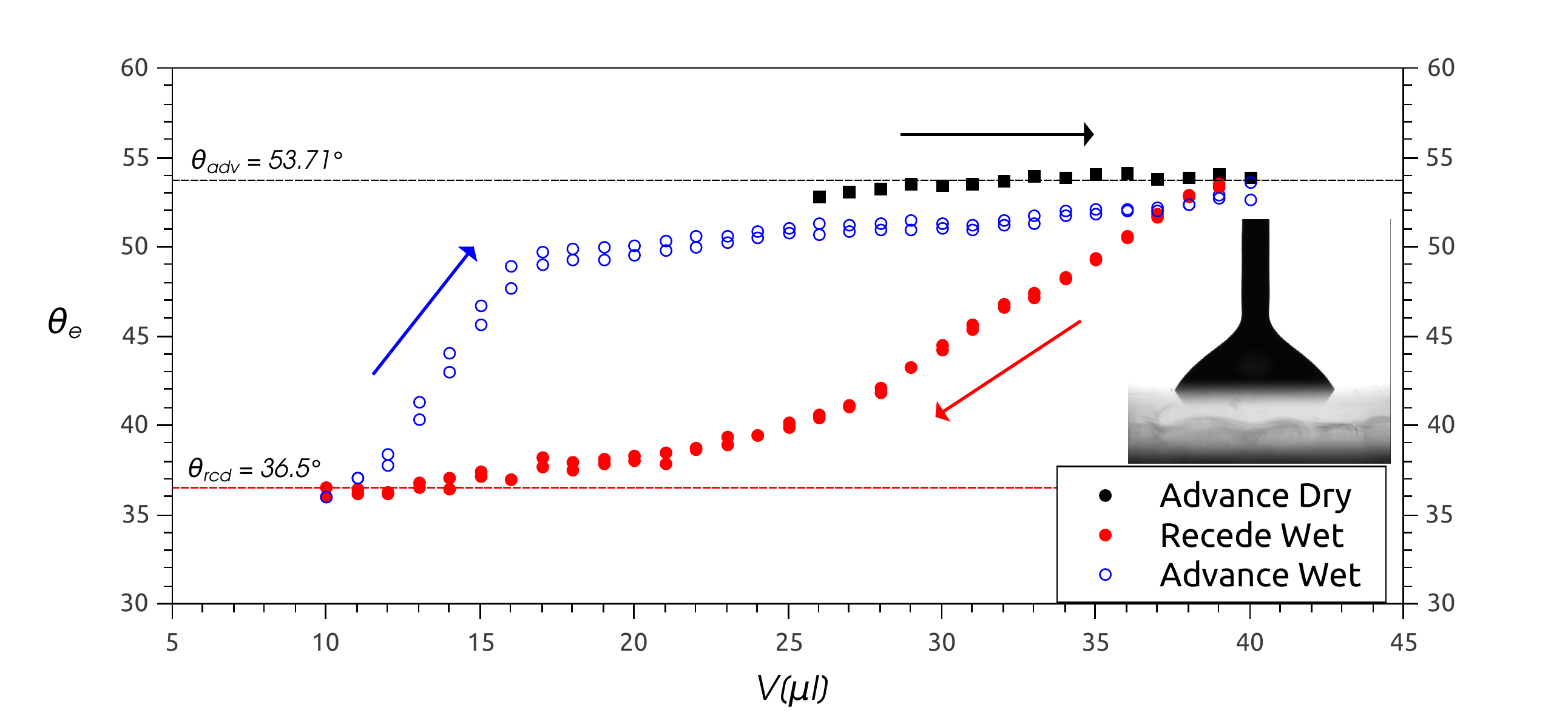}
\caption{Hysteresis cycle of the static contact angle, $\theta_e$. The drop initially spreads on a dry surface (black filled squares) with an increasing volume, and then recedes due to liquid withdrawal (red filled circles). Finally, due to liquid injection, it advances on a prewetted surface (blue hollow circles). The horizontal lines show the maximum and minimum contact angles $\theta_{adv}$ and $\theta_{rcd}$, respectively. The inset shows the drop profile together with the needle for fluid injection/withdrawal.}
\label{fig:hystCycle}
\end{figure}

The wettability of the PDMS on the coated glass was characterized first by measuring the static contact angle, $\theta_e$, of a single sessile circular drop at rest on a horizontal substrate. Then, we were able to quantify the hysteresis cycle of $\theta_e$, as shown in Fig.~\ref{fig:hystCycle}, by using a Rame--Hart Model 250 goniometer. In order to do this, we measure the contact angle when the static state is reached either from a previously advancing or receding contact line. These scenarios are obtained by injecting or withdrawing liquid into an initially circular sessile drop in a controlled manner through a vertical needle that touches the drop at its apex (see inset in Fig.~\ref{fig:hystCycle}). Our system allows to vary the drop volume, $V$, by incremental steps $\Delta \,V = 1\mu l$. 

The goniometer allows to measure $\theta_e$ by analyzing a side view of the drop with a software based on an axisymmetric drop shape technique called ADSA~\cite{kwok_lang97} which fits the front region of the thickness profile by an arc of a circle and allows to have an independent value of the contact angle at each side of the side view of the drop (see inset in Fig.~\ref{fig:hystCycle}). These two values serve as a control of the drop symmetry since the difference between them is less than $1.5 \%$ which is consistent with a circular footprint. We use the average of both values to determine $\theta_e$. This method allows to construct the complete hysteresis cycle shown in Fig.~\ref{fig:hystCycle}. Note that the first branch of the cycle (black solid squares in Fig.~\ref{fig:hystCycle}) is not coincident with the last one (blue hollow circles) for the same volume interval. This is because the initial spreading stage occurs over a dry surface, while the last one (after the dewetting stage) occurs over a surface that has already been pre-wetted the oil (see e.g.~\cite{hennig_04}). It is know that the receding stage does not completely remove all the liquid, but leaves behind a nanometric thin film~\cite{deGennesRMP,beaglehole_jpc89}. Though the difference between $\theta_e$ in both branches is small, we always obtain larger angles for the spreading on a dry surface. In summary, we can characterize the wettability of circular drops by the maximum (minimum) value of the advancing (receding) contact angles of the hysteresis cycle, namely $\theta_{adv}=53.7^\circ$ ($\theta_{rcd}=36.5^\circ$).

\subsection{Liquid filament and drops}
The drops that we analyze in this section are the result of the breakup of a liquid filament placed on the substrate as explained in detail in previous works~\cite{gonzalez_07,gonzalez_04}. We start with a vertical jet of PDMS flowing out from a small nozzle at the bottom of a vessel filled with the silicon oil. The filament is captured from the jet on the substrate by performing suitable rotations of its frame before reaching the final horizontal position. All these movements take about $1-2$ seconds, which is a very short time interval compared to the time scale of the experiment. This procedure yields a fluid filament of uniform width, $w$, with parallel and straight contact lines, so that the initial configuration has a constant cross sectional area along its axis. We calibrate the system by relating the fluid height in the vessel with the jet diameter a given nozzle. Thus, both the jet diameter and the corresponding filament widths (for a contact angle equal to $\theta_{adv}$) could be varied from $0.3$ to $1.3~mm$ and from $0.1$ to $1.0~mm$, respectively. 

\begin{figure}
{\includegraphics[width=0.7\linewidth]{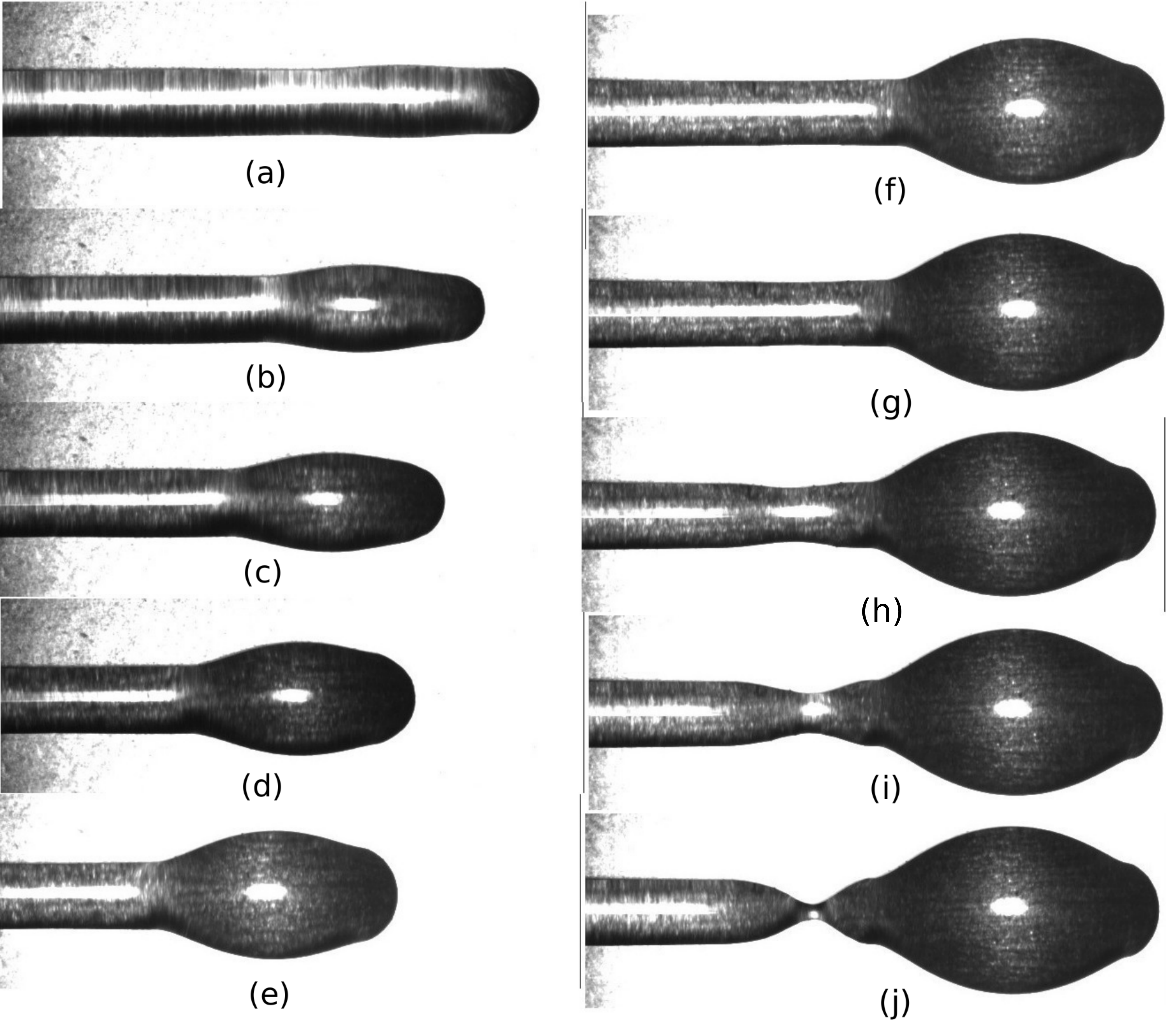}}
\caption{Time evolution of one end of the liquid filament of width $w=0.74~mm$. (a)--(g): The initially rounded extreme recedes and forms a bulge. (h)--(j): A neck develops at a certain distance from the bulge. Finally, this necks breaks up.}
\label{fig:evolFilam}
\end{figure}

A typical breakup process of the filament starts at its extremes, as shown in Fig.~\ref{fig:evolFilam} for one extreme. Initially, the ends are rounded and recede. Then, the receding flow forms a bulge, Fig.~\ref{fig:evolFilam}(a)-(g), and a neck develops at a certain distance from it, Fig.~\ref{fig:evolFilam}(h)-(j). Finally, this neck breaks up, and the remaining bulge completes its dewetting process by evolving into a sessile drop with non--circular footprint. This mechanism repeats itself starting from the new end of the filament, and finally gives place to a series of similar drops. Here, we are interested mainly on the shape of the drop that results from this evolution, and not in the description of the instability itself, which has been thoroughly studied elsewhere~\cite{diez_04,gonzalez_07,dgk_pof09,dk_JPCS09}.

\begin{figure}[htb]
{\includegraphics[width=0.7\linewidth]{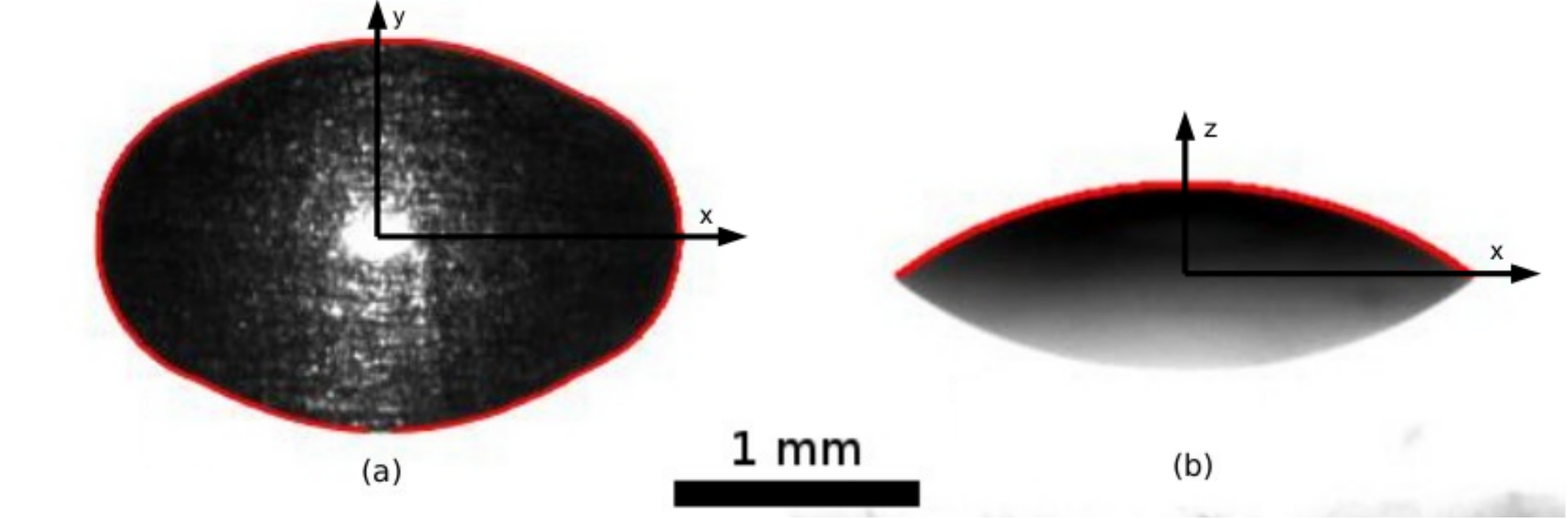}}
\caption{ (a) Top, and (b) side views of a drop. The red lines are the digitally obtained lines for the footprint and the longitudinal thickness profile, respectively. The dimensional widths in $x$ and $y$ directions are $2.362~mm$ and $1.588~mm$, respectively, the maximum thickness at the drop center is $0.383~mm$, and the estimated drop volume is $\bar V= 7.75 \mu l$.}
\label{fig:CartFootprint}
\end{figure}

In Fig.~\ref{fig:CartFootprint} we show the top and side views of a drop together with the digital curves (red lines) obtained by image analysis. The side view in Fig.~\ref{fig:CartFootprint}(b) corresponds to the longitudinal direction (i.e. along the axis of the original filament, $x$--axis). Unfortunately, the transverse view (along $y$--axis) cannot be obtained since it is blocked by similar drops sit along the same line. The maximum and minimum diameters of the footprint are $w_x=1.628$ and $w_y=1.095$, while the maximum thickness is $h_{max}=0.264$ (all variables in units of $a_c$).  

Since the drop is clearly non axisymmetric, its volume cannot be determined by the ADSA method. Instead, the average volume, $\bar V$, of the drops resulting from a given filament can be estimated from the number of drops, the weight of the substrate with the drops, and the liquid volume of the jet that formed the original filament. In our experiment, we have $\bar V=0.254$ in units of $a_c^3$. This parameter is not needed for the following analytical modeling, but is necessary to perform the simulations in Section~\ref{sec:num}.

\begin{figure}
\centering
\subfigure []
{\includegraphics[width=0.3\linewidth]{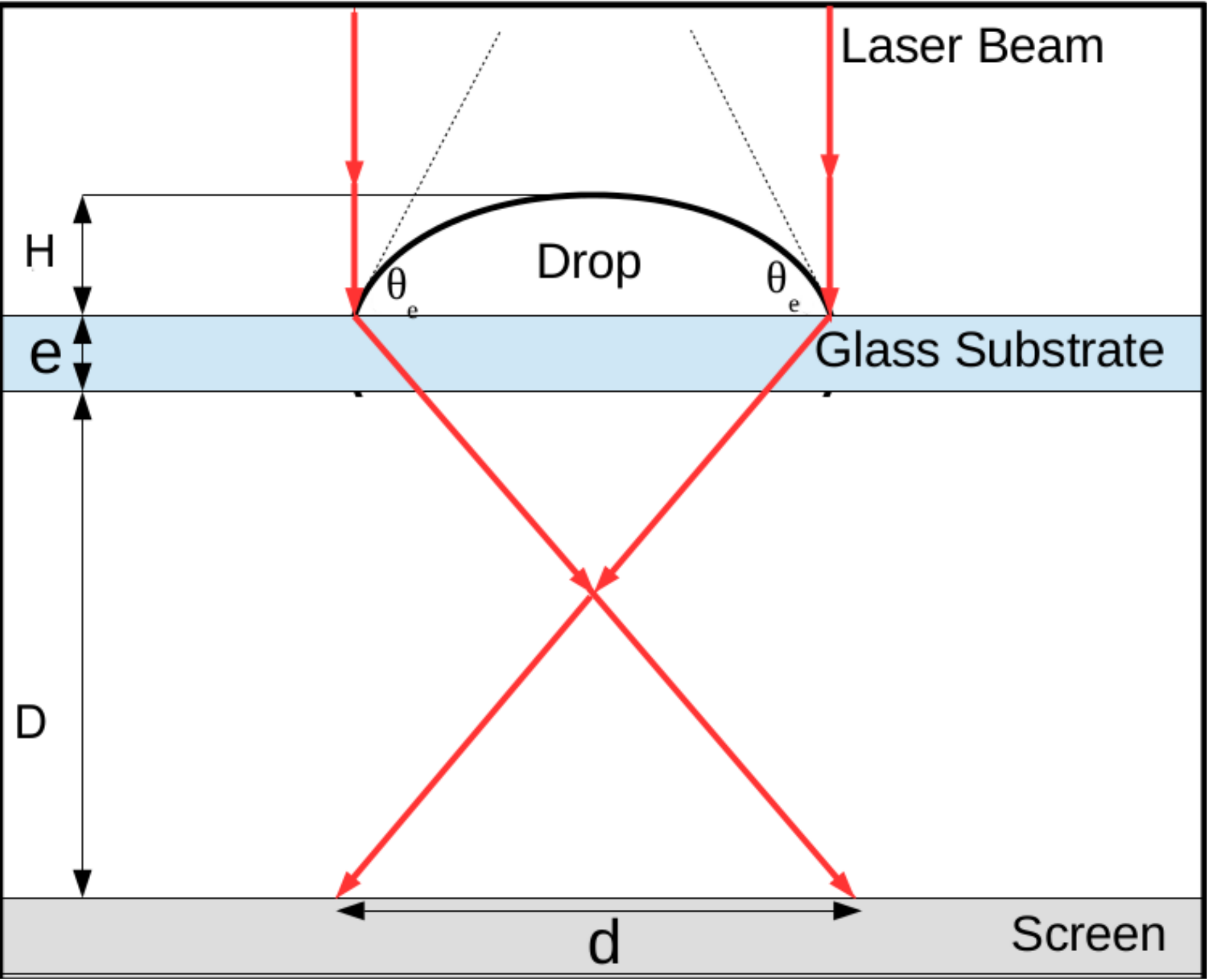}}\hspace{2cm}
\subfigure []
{\includegraphics[width=0.2\linewidth]{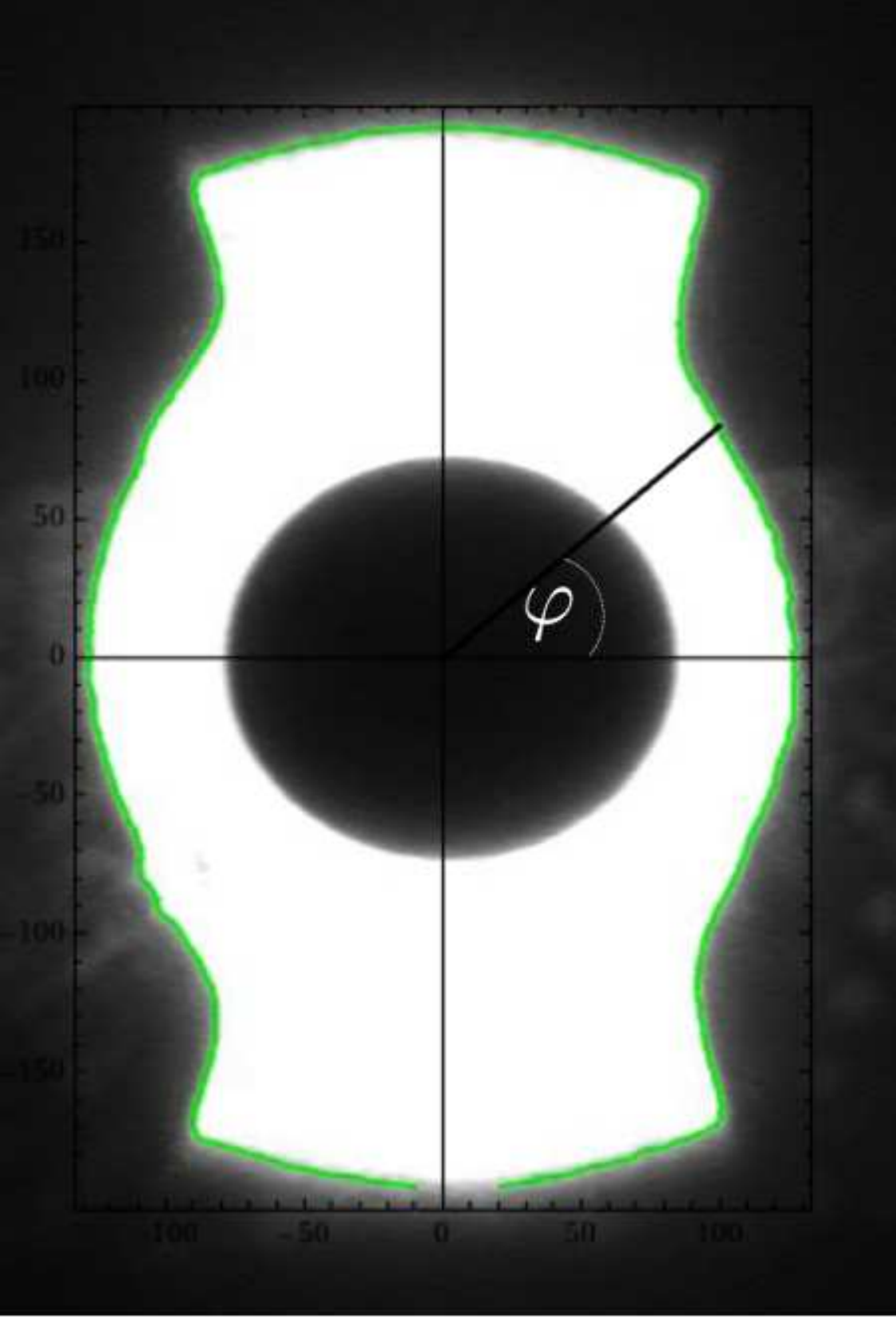}}
\caption{(a) Light rays diagram of the refractive method. (b) Refraction pattern of the drop in Fig.~\ref{fig:CartFootprint} as seen on the screen. The green line shows the digital capture of the image contour.}
\label{fig:refrac}
\end{figure}

Since the standard technique of the goniometer can measure contact angles only on vertical planes along the axes of symmetry, for a drop with non--circular footprint we must resort to a different technique. Here, we use one that is based on the refraction of light by the contact line region~\cite{gonzalez_04}. Thus, we impinge the drop from above with an He--Ne laser beam, normally to the substrate. This beam, which is wider than the footprint, is refracted by the drop free surface and produces an illuminated area on a screen placed at a distance, $D \approx 5~cm$, below the substrate. For a footprint of arbitrary shape (non--circular), the maximum light deflection angle, $\beta(\varphi)$, varies at different azimuthal directions, $\varphi$, with  $0\leq \varphi \leq 2\pi$.  Thus, it is given by $\tan \beta =d/(2D)$, where $d$ is the diameter of the illuminated figure at a certain $\varphi$, which passes through the drop center. Then, the contact angle is given by
\begin{equation}
 \tan \theta_e= \frac{\sin \beta}{\sqrt{n^2 - \sin^2 \beta }-1}
 \label{eq:theta_e}
\end{equation}
where $n = 1.4$ is the refraction index of PDMS. This expression assumes that the glass and drop thicknesses, $e$ and $H$, respectively, are much smaller than the distance to the screen, $D$, see Fig.~\ref{fig:refrac}(a). Thus, Eq.~(\ref{eq:theta_e}) allows to obtain the contact angle of the drop, $\theta_e(\varphi)$, around all of its periphery. A typical illumination pattern is shown in Fig.~\ref{fig:refrac}(b), where the central region is blocked with a light stopper to avoid saturation of the camera in order to obtain a good contrast. The diameter of this stopper, which is actually a coin, also works as a reference of length scale. 

Interestingly, the contact angles given by Eq.~(\ref{eq:theta_e}) in the transverse and longitudinal directions ($\varphi=0$ and $\pi/2$ in Fig.~\ref{fig:refrac}(b)) are very close to maximum (minimum) value of the advancing (receding) contact angles measured in the hysteresis cycle, namely $\theta_{adv}$ ($\theta_{rcd}$) in Fig~\ref{fig:hystCycle}. This is because the drop adopts its final shape by wetting in the transverse direction and dewetting in the longitudinal one. Therefore, these two angles characterize the resulting geometry of the footprint.

\section{Formalism for the non--circular drop}
\label{sec:form}

We are now concerned with static solutions of the dimensionless Navier--Stokes (N-S) equation,
\begin{equation}
La \left[ \frac{\partial \vec {v}}{\partial t}+(\vec{v} \cdot \vec {\nabla} ) \cdot \vec{v} \right] 
= - \vec {\nabla} p + \nabla ^{2} \vec {v} -\vec{z},
\label{eq:NS}
\end{equation}
where the last term stands for the gravity force. Here, the scales for the position $\vec x=(x,y,z)$, time $t$, velocity $\vec {v}$ and pressure $p$ are $a_c$, $\mu a_c/\gamma$, $\gamma/\mu$, and $\gamma/a_c$, respectively. Therefore, the Laplace number is $La=\rho \gamma a_c/\mu^2$. 
The $x$ and $y$--axes are assigned along and across the original filament, respectively.

For a sessile drop at rest, we have $\vec v=0$, so that the balance between gravity and pressure gradient is given by, 
\begin{equation}
0 = -\vec{z} - \vec {\nabla} p,  
\end{equation}
plus the boundary condition of capillary pressure jump at the interface: 
\begin{equation}
 p_0-p(z=h)=p_{cap}=\frac {\nabla^2 h}{\left( 1+ |\nabla h |^2 \right)^{3/2}},
 \label{eq:pcap}
\end{equation}
where $h(x,y)$ is the drop thickness function, and $p_0$ is the ambient pressure outside the drop. If follows that the pressure field in the sessile drop is $p(z)=p_0-z-p_{cap}$. Since the equilibrium condition is $p=const.$, we have 
\begin{equation}
 p_{cap}+h=p_1.
\label{eq:p_eq}
\end{equation}
where $p_1$ is the (unknown) constant pressure inside the drop. 

When the drop is axisymmetric, i.e. with circular footprint, the static problem posed by Eqs.~(\ref{eq:pcap}) and (\ref{eq:p_eq}) is already solved in the literature (see e.g. \cite{michael_prsl77}). However, no similar attempts have been done for the non--axisymmetric problem. Here, we retain the asymmetric features of the drop, but somewhat simplify the equations by assuming the validity of the lubrication theory. Within this approach, the long wave approximation requires small slopes at the free surface, so that $|\nabla h | \ll 1$, and Eq.~(\ref{eq:p_eq}) becomes
\begin{equation}
 \nabla^2 h + h =p_1,
\label{eq:p_eq1}
\end{equation}
which is the inhomogeneous Helmholtz equation. The boundary conditions are $h=0$ and given contact angle, $\partial h /\partial \hat {\bf n}$, along the contact line. A solution of this equation can be written in the form $h= p_1 + h_1$, where $h_1$ is the solution of the homogeneous equation, 
\begin{equation}
 \nabla^2 h_1 + h_1 =0.
 \label{eq:ode_h1}
\end{equation}
 
Our approach in this paper is to compare particular solutions of Eq.~(\ref{eq:p_eq1}) with experimental data. These solutions have been selected as having a special symmetry and requiring the smallest number of parameters from the experiments. This comparison is performed by verifying that the predicted shape of the contact line, thickness profiles, and contact angles are in agreement with the measurements.

\section{Static solution in Cartesian coordinates}
\label{sec:lub_cart}

Since the experimental data show that the characteristic contact angles, $\theta_{adv}$ and $\theta_{rcd}$, correspond to perpendicular axes, we consider the solution as having two axes of symmetry, namely $x$ and $y$. Therefore, expressing the solution is Cartesian coordinates and assuming variable separation, we write $h_1= X(x) Y(y)$. Then, Eq.~(\ref{eq:ode_h1}) leads to
\begin{equation}
\frac{X^{\prime\prime}}{X}+\frac{Y^{\prime\prime}}{Y}+1=0.
\end{equation}
Among all possible solutions of this equation, we restrict ourselves to even functions in both $x$ and $y$, due to the symmetric character of the drop respect to $x$ and $y$ reflections. Thus, we find
\begin{equation}
h_1=-A \cosh a x \cosh b y,
\label{eq:h1A}
\end{equation}
where $A$ is a constant, and $a,b$ are determined by the constraint
\begin{equation}
a^2+b^2=1.
\label{eq:ab1}
\end{equation}
Note that, in general, the solution is an integral in terms of $a$ with $A(a)$. However, for simplicity, we assume that a single value of $a$ is needed to reasonably describe the solution. This is in consonance with our aim to determine the detailed properties of the experiments from solutions that comply with only a small subset of the boundary conditions. This allows us to test the robustness of the solutions obtained from this model.

In general, a boundary condition is $h(x_c,y_c)=0$ at the points $(x_c,y_c)$ of the contour of the footprint (contact line), which means that 
\begin{equation}
p_1=-h_1(x_c,y_c). 
\end{equation}
Then, the widths (or diameters) of the drop along the $x$ and $y$-axes, namely $w_x$ and $w_y$, can be obtained for $y_c=0$ and $x_c=0$, respectively. Thus, the  solution of the type~(\ref{eq:h1A}) yields 
\begin{equation}
p_1=A \cosh b w_y, \qquad  p_1=A \cosh a w_x,
\label{eq:pvsA}
\end{equation}
which implies
\begin{equation}
\frac{b}{a}=\frac{w_x}{w_y}.
\label{eq:bavsw}
\end{equation}
Since the contact angles at the axes of symmetry of the drop are given by
\begin{eqnarray}
\theta_x&\approx& -\left.\frac{\partial h}{\partial x}\right|_{y=0}= -a A \sinh a w_x,
\label{eq:angr} \\
\theta_y &\approx& -\left.\frac{\partial h}{\partial y}\right|_{x=0}= -b A \sinh b w_y,
\label{eq:anga}
\end{eqnarray}
we obtain the relationship
\begin{equation}
\frac{\theta_x}{\theta_y}=\frac{w_y}{w_x}.
\label{eq:thetavsw}
\end{equation}

\begin{figure}
\includegraphics[width=0.5\textwidth]{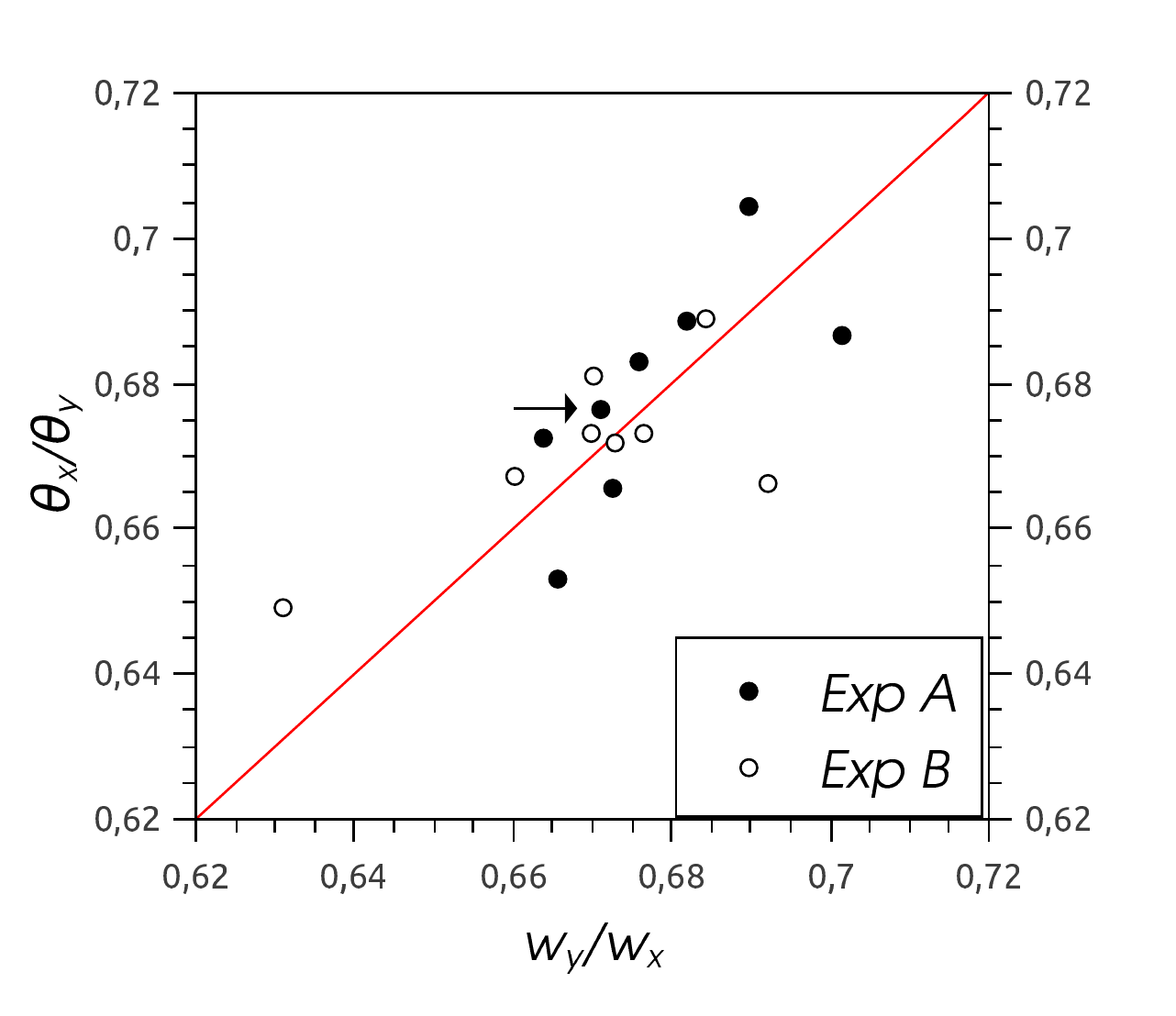}
\caption{Ratios between contact angles and diameters of drops from two filaments, A and B, of the same length and width. The straight line corresponds to Eq.~(\ref{eq:thetavsw}), and the arrow points the datum corresponding to the drop in Fig.~\ref{fig:CartFootprint}}
\label{fig:lin_rel}
\end{figure}

Although this equation is obtained within the long wave approximation where small slopes are assumed, we test its range of validity by comparing it with  experimental data corresponding to relatively large angles. Figure~\ref{fig:lin_rel} shows this comparison, where the ratios for drops belongs to two filaments, A and B, of the same length and width. Even after taking into account the dispersion of the data, one can conclude that they basically follow the linear trend with slope equal to one as predicted by Eq.~(\ref{eq:thetavsw}). 

Note that the maximum thickness of the drop, $h_{max}$, occurs at $(x,y)=(0,0)$, i.e the center of the drop. At this point, the following relation holds
\begin{equation}
h_{max}=p_1-A.
\label{eq:hmax}
\end{equation}

\begin{figure}[htb]
\subfigure []
{\includegraphics[width=0.3\linewidth]{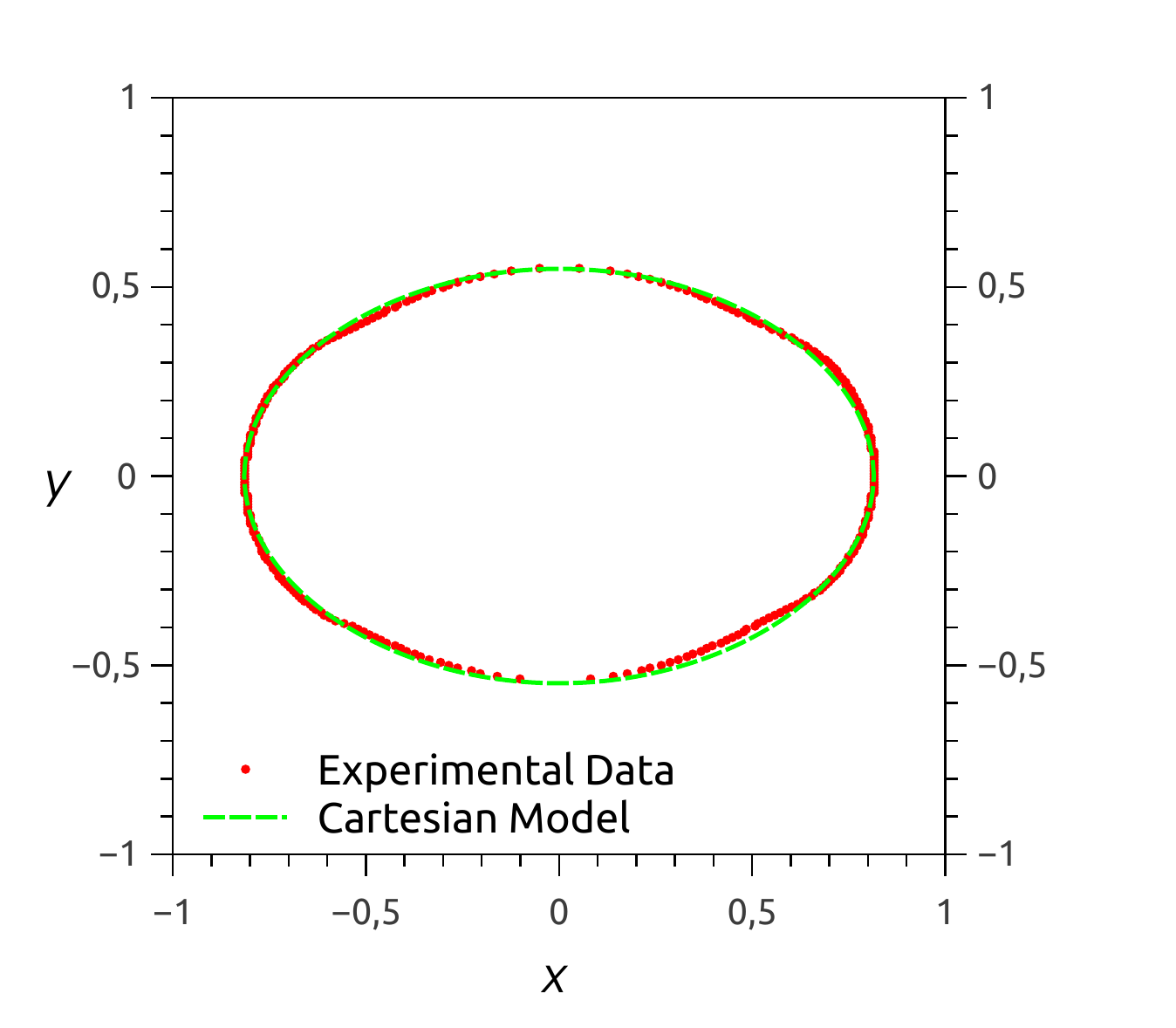}
\label{FootCartExp}}
\subfigure []
{\includegraphics[width=0.3\linewidth]{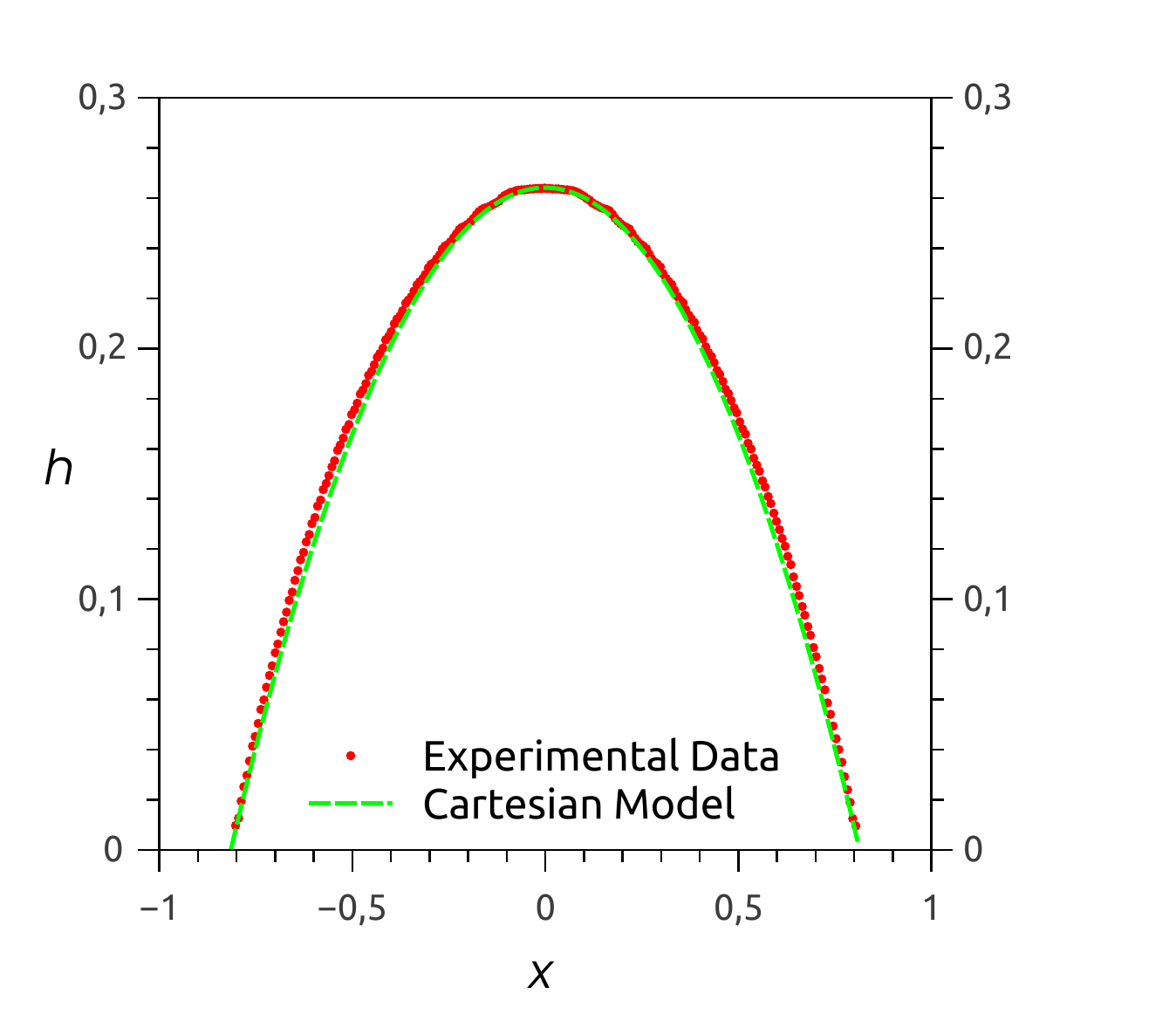}
\label{hxCartExp}}
\subfigure []
{\includegraphics[width=0.3\linewidth]{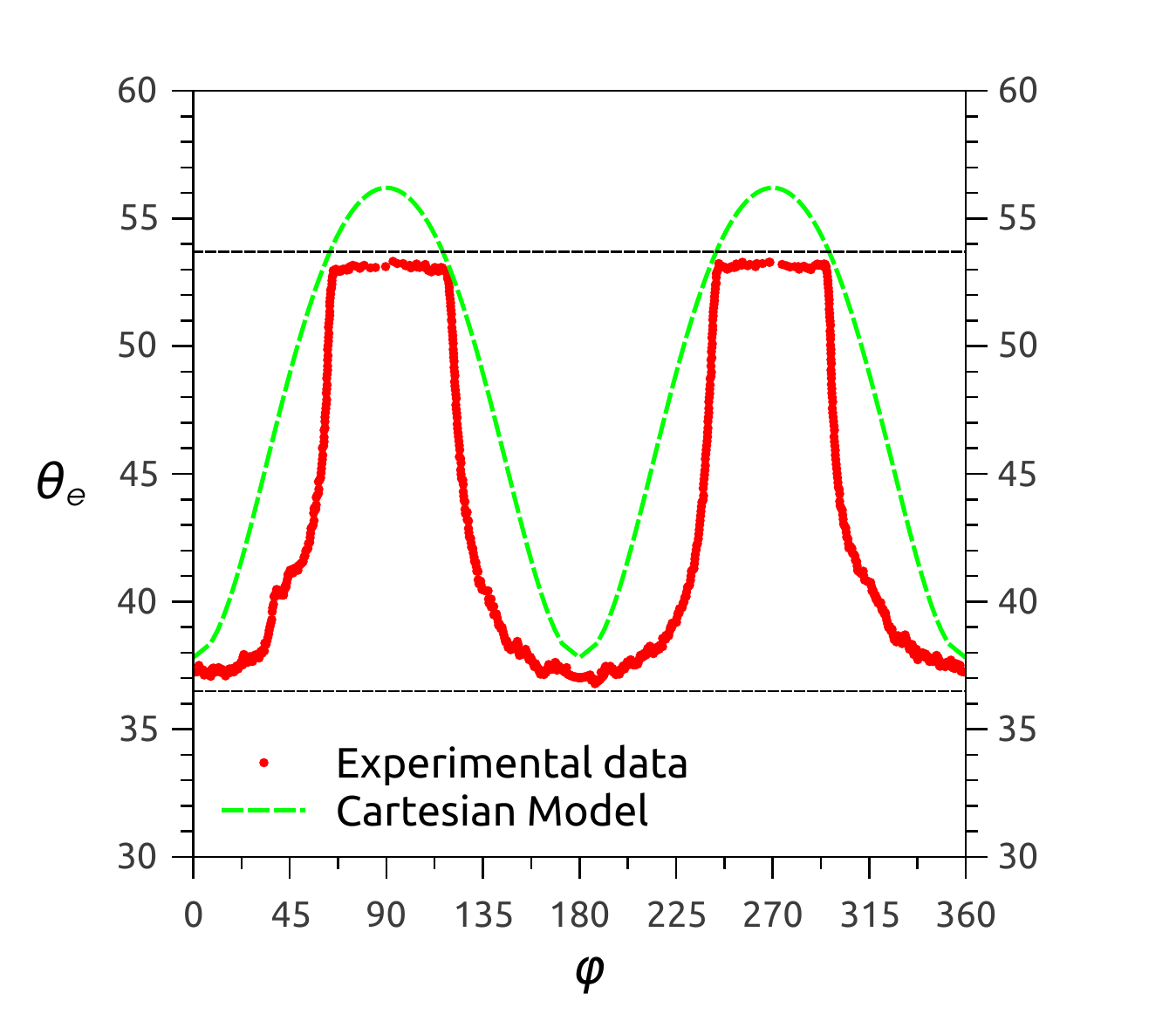}
\label{AngCartExp}}
\caption{Comparison between the experimental data (red points) and the Cartesian solution (green dashed lines) using $w_x$, $w_y$, and $h_{max}$ as input data for: (a) drop footprint, (b) longitudinal thickness profile along the filament ($x$--axis), and (c) contact angle distribution around the footprint periphery. The horizontal dashed lines correspond to $\theta_{rcd}$ and $\theta_{adv}$ shown in Fig.~\ref{fig:hystCycle}.}
\label{fig:Exp_cart}
\end{figure}

One special feature of this Cartesian solution is that it is determined by assigning the values of three parameters. For example, one could choose the group: $w_x$, $w_y$, and $h_{max}$. For this case, we show in Fig.~\ref{fig:Exp_cart} the comparison between the theoretical results and the experiment for the shape of the footprint, the longitudinal thickness profile, and the contact angle distribution around the drop periphery. The theoretical solution is obtained by using Eqs.~(\ref{eq:ab1}) and (\ref{eq:bavsw}) to find $a$ and $b$, and Eq.~(\ref{eq:pvsA}) to obtain the ratio $p_1/A$. Finally, by introducing the result into Eq.~(\ref{eq:hmax}), we have $p_1$ for given $h_{max}$.  Note that, aside from the thickness at the drop center, $h_{max}$, this solution is determined by using information only from two points of the contact line, which are the measured values of the widths at the symmetry axes. Interestingly, Figs.~\ref{FootCartExp} and \ref{hxCartExp} show that the footprint as well as the thickness profile present only very small departures from the experimental data. However, the differences in the azimuthal distribution of contact angles are important at the maximums and we find a much smoother transition regions between the maximum and minimum contact angles (see  Fig.~\ref{AngCartExp}), likely because the three parameters used in this solution do not contain sufficient angular information. 

\begin{figure}[htb]
\subfigure []
{\includegraphics[width=0.3\linewidth]{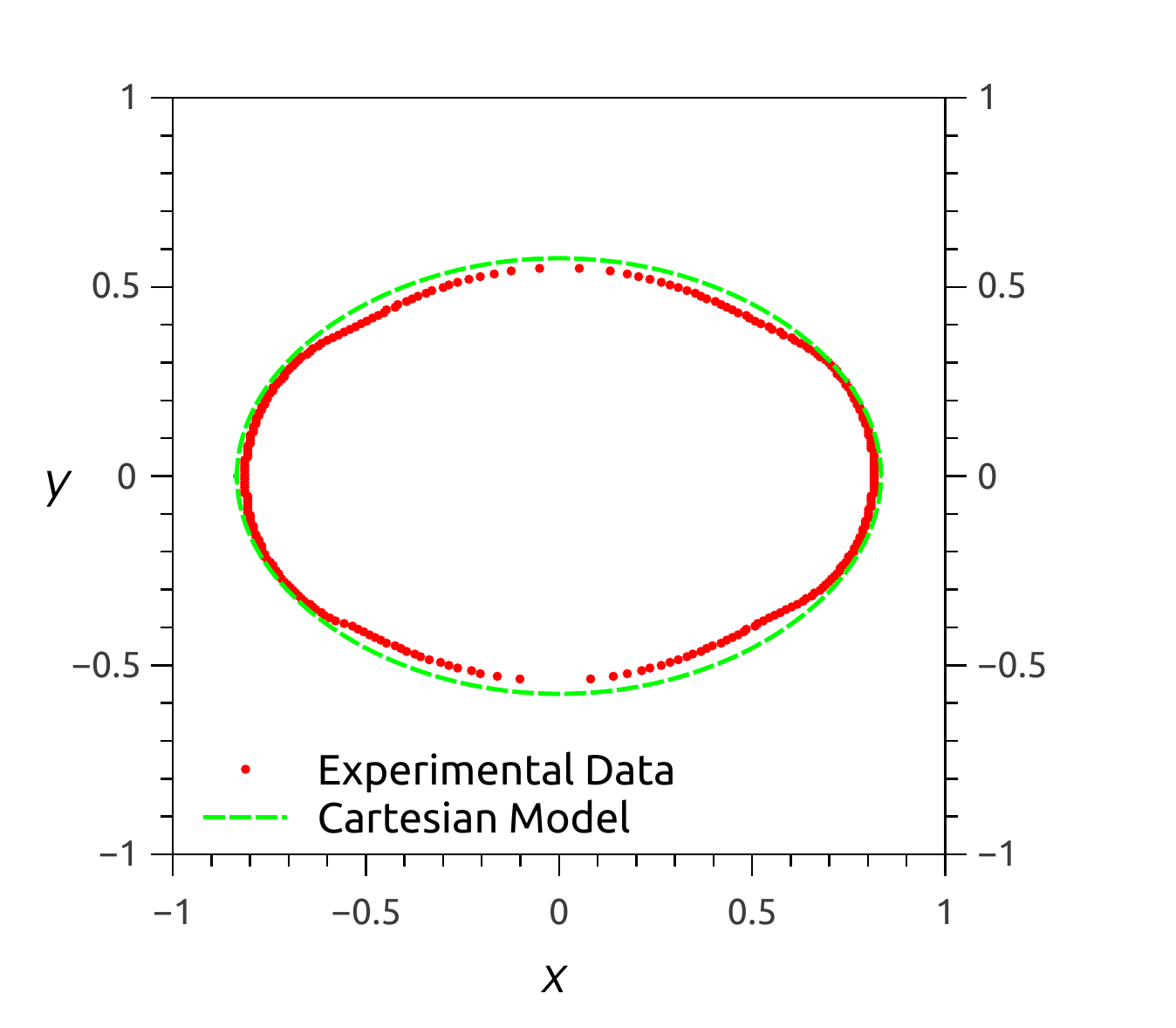}
\label{FootCartExpAng}}
\subfigure []
{\includegraphics[width=0.3\linewidth]{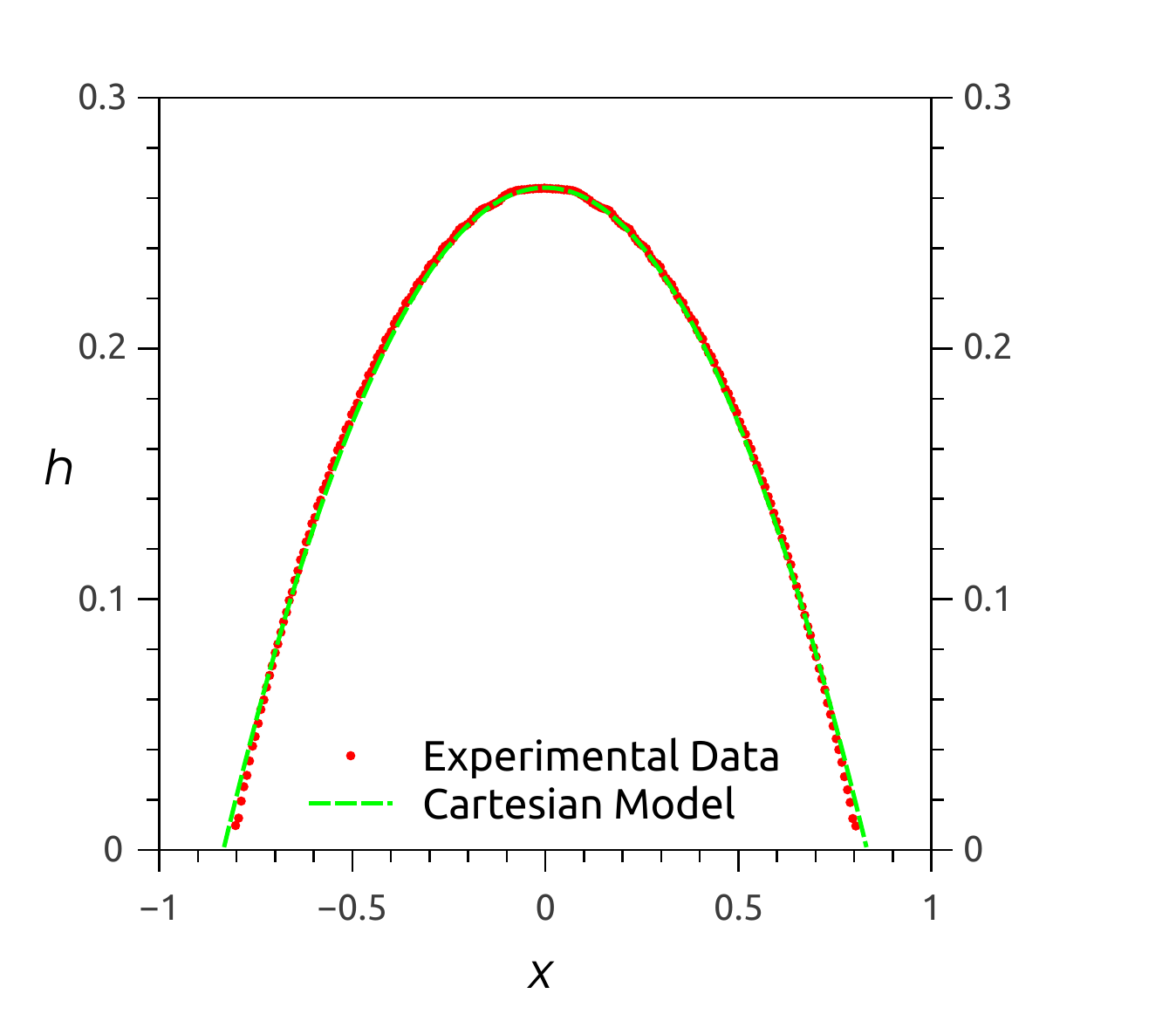}
\label{hxCartExpAng}}
\subfigure []
{\includegraphics[width=0.3\linewidth]{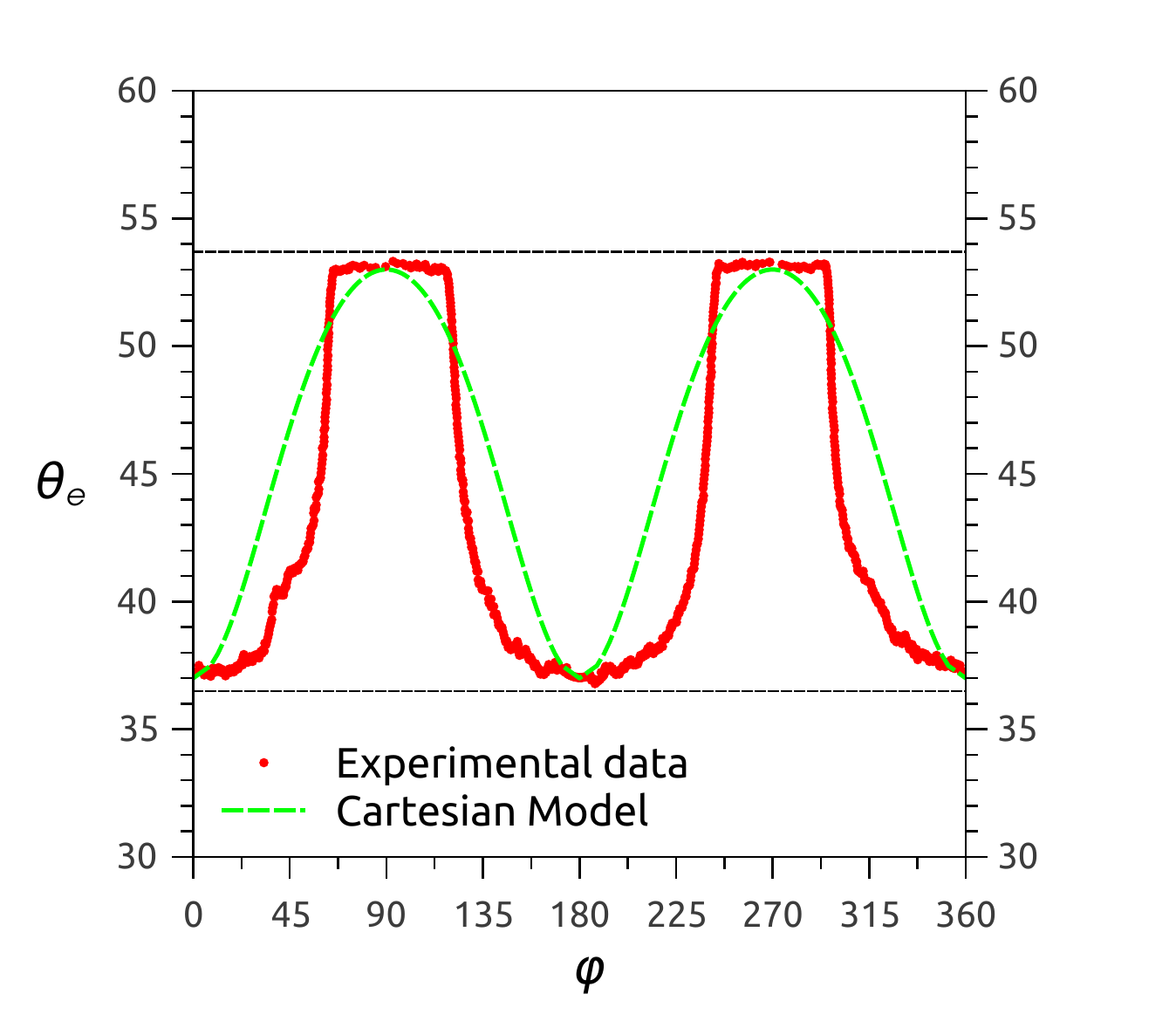}
\label{AngCartExpAng}}
\caption{Comparison between the experimental data (red points) and the Cartesian solution (green dashed lines) using $\theta_x$, $\theta_y$, and $h_{max}$ as input data for: (a) drop footprint, (b) longitudinal thickness profile along the filament ($x$--axis), and (c) contact angle distribution around the footprint periphery. The horizontal dashed lines correspond to $\theta_{rcd}$ and $\theta_{adv}$ shown in Fig.~\ref{fig:hystCycle}.}
\label{fig:Exp_cart_ang}
\end{figure}

On the other hand, another example could be given by using $\theta_x$ and $\theta_y$ instead of $w_x$ and $w_y$ as input data (aside from $h_{max}$). Now, the results look a bit different due to the inclusion of angular information (see Fig.~\ref{fig:Exp_cart_ang}). In this case, we obtain a better comparison for the contact angle distribution, Fig.~\ref{AngCartExpAng}, along with some differences in the shape of the footprint, Fig.~\ref{FootCartExpAng}, but no visible departures at the thickness profile, Fig.~\ref{hxCartExpAng}. Naturally, other combinations of three parameters could be used, but the above cases are enough to show the sensitivity of the results respect to the choice. Note, in passing, that the ratio $w_y/w_x$ for this drop is in good agreement with Eq.~(\ref{eq:thetavsw}) (see arrow in Fig.~\ref{fig:lin_rel}). 
In summary, in spite of this small number of parameters, the predictions are remarkably good for the shape and thickness, while the agreement for the contact angle distribution is less satisfactory. The fact that this Cartesian solution fails to fully describe the azimuthal distribution of the contact angle (see  Fig.~\ref{AngCartExp}) implies that more information is needed to describe the static free surface.  Note that this Cartesian model is not completely general since it is limited to only two symmetries, namely respect to the $x$ and $y$--axes. Moreover, it does not fully take into account any harmonic component of the problem, other than angular periodicities multiples of $\pi/2$. In order to improve this solution, we should resort to an integral, which requires a not straightforward procedure. 

\section{Static solution in polar coordinates}
\label{sec:lub_polar}

Instead of increasing the complexity of the Cartesian model, we show here that polar coordinates are more convenient to deal with other details of the problem.  They allow us to apply the basic angular symmetries of the drop in a very convenient way and reduce the problem to a Fourier series instead of an implicit integral equation.

If the problem is expressed in the radial and angular variables $(\rho,\varphi)$, a separation of the form
\begin{equation}
h_1=R(\rho)\Phi(\varphi)
\end{equation}
is possible, leading to two uncoupled ordinary differential equations for $R$ and $\varphi$ from Eq.~(\ref{eq:ode_h1}). Note that since we must assume angular periodicity in $0\leq \varphi \leq 2\pi$, the solution in $\varphi$ satisfies a Sturm-Liouville problem and a Fourier series is obtained. The equation for $h_1$ can be written as
\begin{equation}
	\frac{\rho }{R}\frac{d }{d \rho}\left(\rho \frac{d R}{d \rho}\right) +\frac{1}{ \varphi}\frac{d^2 \Phi}{d \varphi^2}+\rho^2=0,
\end{equation}
where the $\varphi$--dependence is of the form
\begin{equation}
\varphi= A_m \cos m\varphi + B_m \sin m \varphi,
\end{equation} 
with $m$ a positive integer. Thus, it turns out that $R$ is given by the Bessel function of order $m$ (Neumann functions must be discarded since they diverge at the center of the drop). Therefore, the general solution in polar coordinates can be expressed as
\begin{equation}
h(\rho,\varphi)=p_1+ \sum_{m=0}^\infty(A_m \cos m\varphi + B_m \sin m \varphi) J_m(\rho).
\label{eq:hpolar}
\end{equation}

In order to have a symmetric solution respect to reflections along $x$ and $y$--axes, we must have $B_m=0$ and odd $m$. Here, we further assume that the shape of the drop can be reasonably estimated by the first five terms of Eq.~(\ref{eq:hpolar}), so that
\begin{equation}
h(\rho,\varphi)\approx p_1+A_0 J_0(\rho)+A_2 J_2(\rho) \cos 2 \varphi + A_4 J_4(\rho) \cos 4 \varphi + A_6 J_6(\rho) \cos 6 \varphi,
\label{eq:polarh1}
\end{equation}
where the five unknown constants, $(p_1,A_0,A_2,A_4,A_6)$, must be determined from the experimental data. In fact, by measuring the values of $(w_x,w_y,h_{max},\theta_x,\theta_y)$, we can form the following system of independent equations,
\begin{equation} 
 h \left( \frac {w_x}{2},0 \right)=0, \qquad h\left( \frac {w_y}{2},\frac {\pi}{2} \right)=0, \qquad h(0,0)=h_{max},
 \qquad \frac {\partial h}{\partial x}\left( \frac {w_x}{2},0 \right)=\theta_x, \qquad \frac {\partial h}{\partial y}\left( \frac {w_y}{2},\frac {\pi}{2} \right)=\theta_y,
 \label{eq:condic}
\end{equation}
which can be solved analytically.

\begin{figure}[htb]
\subfigure []
{\includegraphics[width=0.3\linewidth]{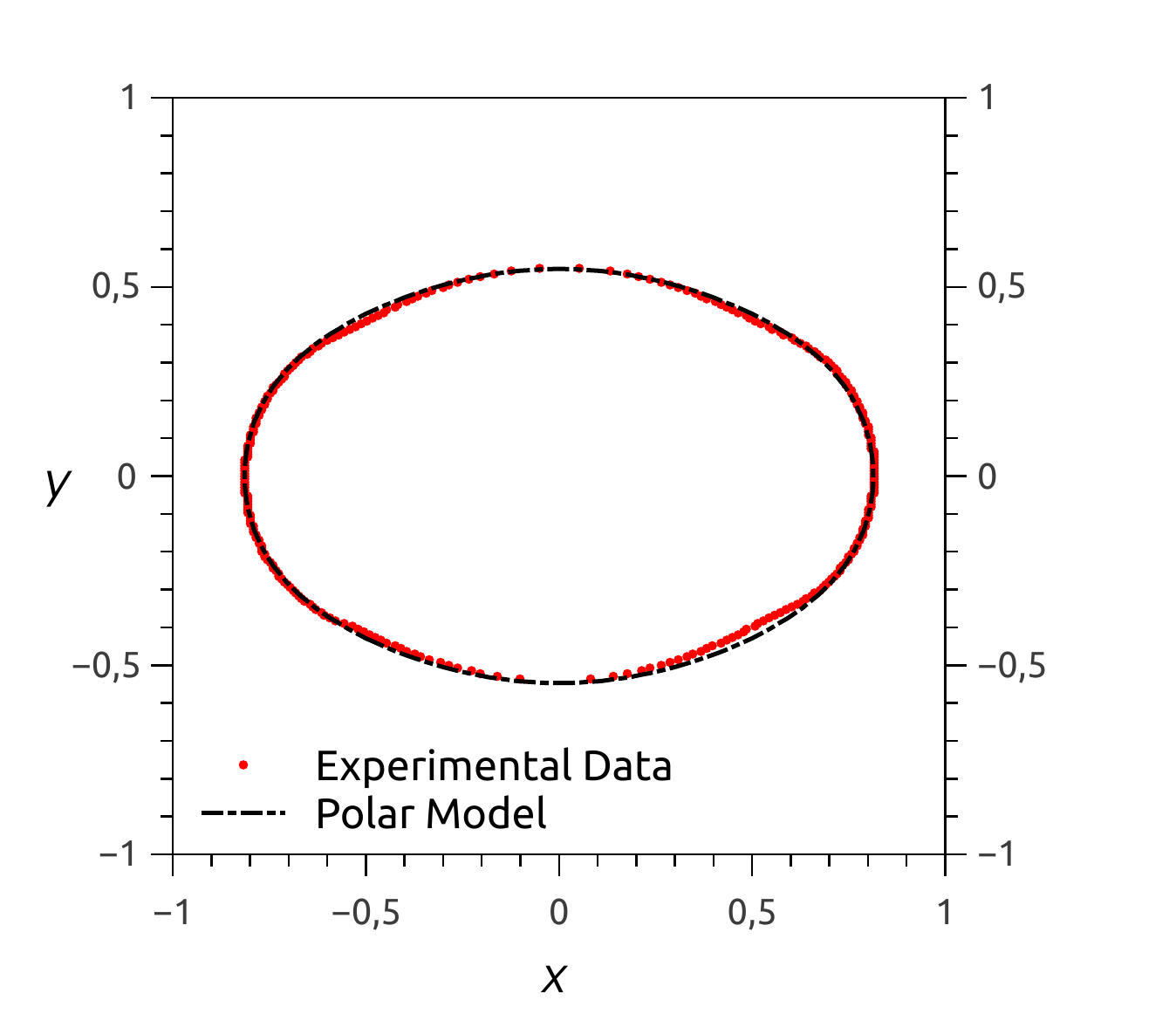}
\label{FootPolarExp}}
\subfigure []
{\includegraphics[width=0.3\linewidth]{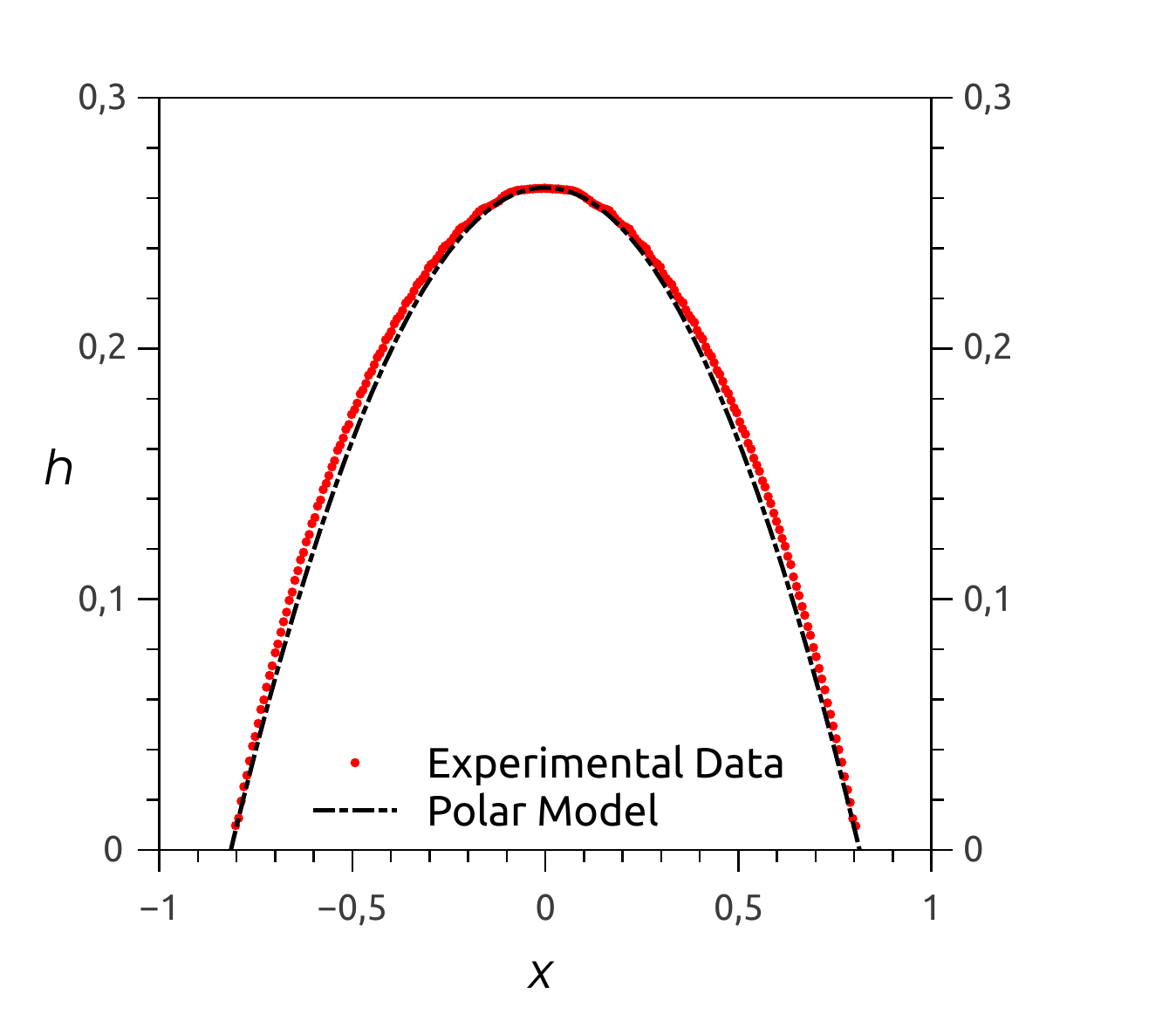}
\label{hxPolarExp}}
\subfigure []
{\includegraphics[width=0.3\linewidth]{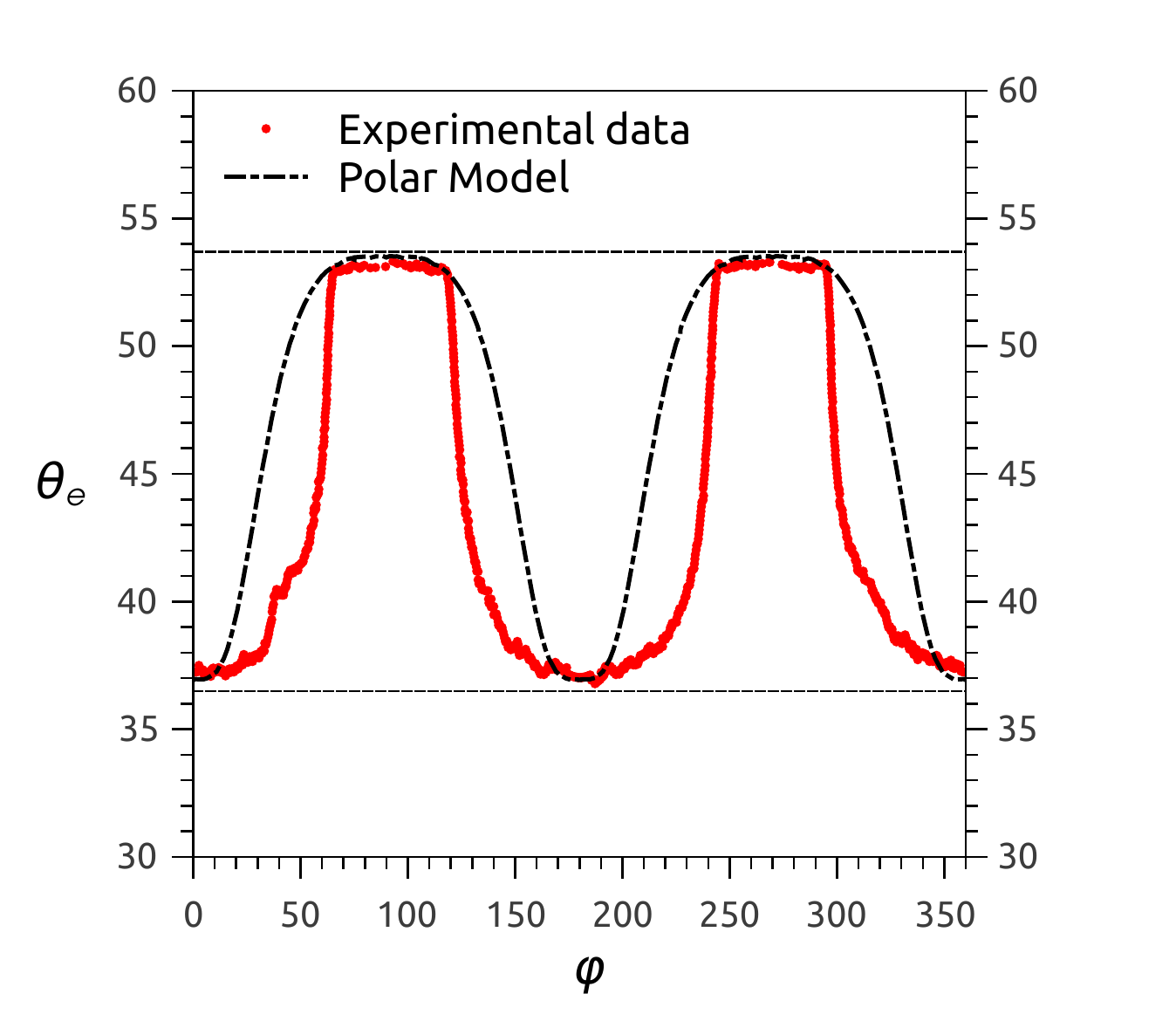}
\label{AngPolarExp}}
\caption{Comparison between the experimental data (red points) and the polar solution (black dashed--dotted lines) determined with \emph {five parameters} for: (a) drop footprint, (b) longitudinal thickness profile along the filament ($x$--axis), and (c) contact angle distribution around the footprint periphery. The horizontal dashed lines correspond to $\theta_{rcd}$ and $\theta_{adv}$ shown in Fig.~\ref{fig:hystCycle}.}
\label{fig:Exp_Polar}
\end{figure}

The comparison between this solution and experiments is shown in Fig.~\ref{fig:Exp_Polar}. Similarly to the Cartesian solution, Figs.~\ref{FootPolarExp} and \ref{hxPolarExp} show that the footprint as well as the thickness profile given by the polar solution present very little difference with the experimental data. However, unlike the Cartesian solution, the polar one clearly gives a much better approximation to the experimental contact angles. In fact, Fig.~\ref{AngPolarExp} shows a very good agreement not only by yielding accurate values of the advancing and receding contact angles, but also in the regions nearby these angles. Nevertheless, even if the transition zones are steeper than the Cartesian solution, they are not still steep enough when compared to the experimental data. We believe that this is because the detailed information about these regions require more than four terms of Eq.~(\ref{eq:polarh1}). In other terms, the small amount of experimental data from Eq.~(\ref{eq:condic}) are not enough to determine the steep change in the contact angle.

\subsection{Selection of minimum number of parameters}
\begin{figure}[ht]
\subfigure []
{\includegraphics[width=0.3\linewidth]{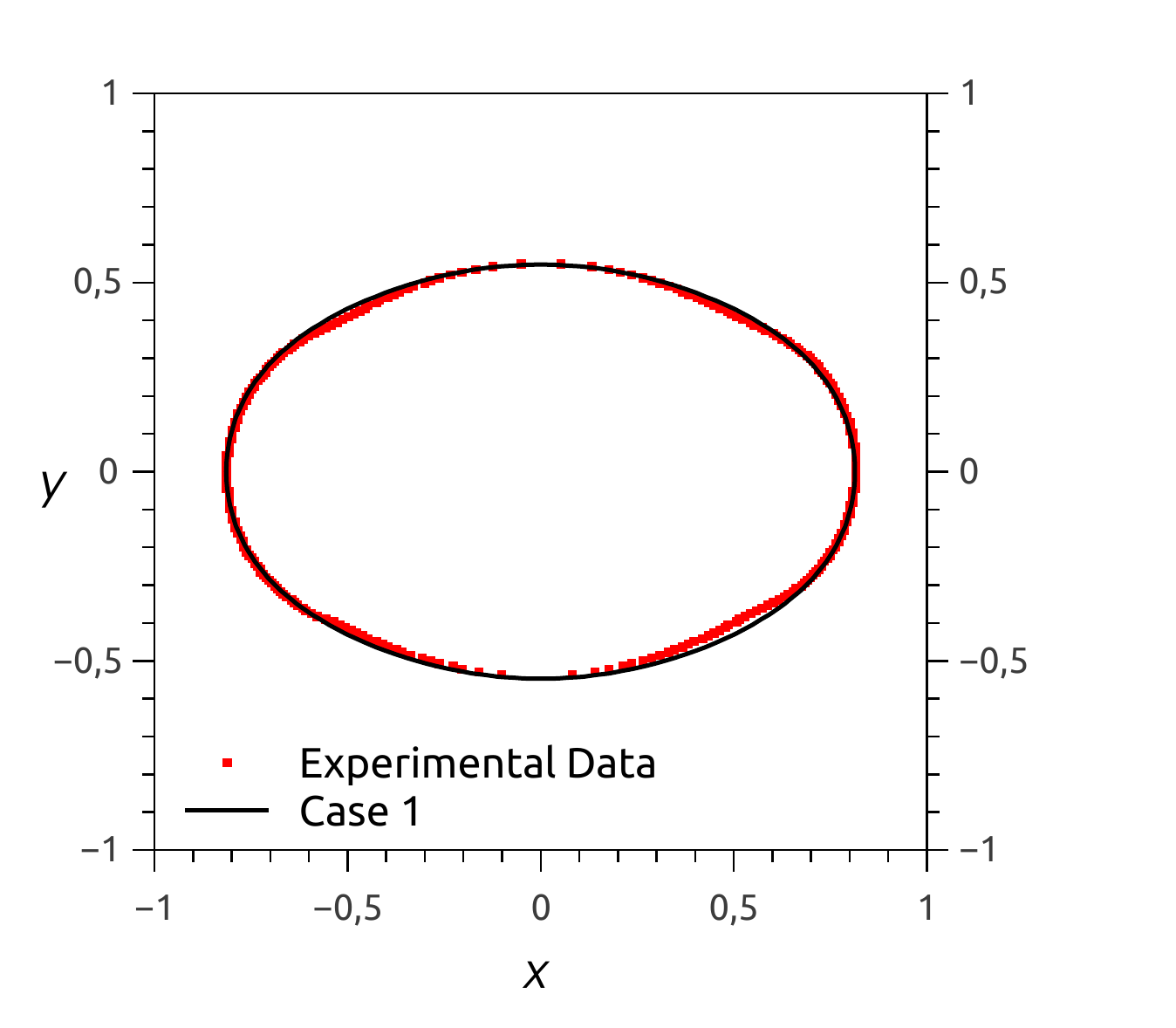}
\label{FootPolarC1}}
\subfigure []
{\includegraphics[width=0.3\linewidth]{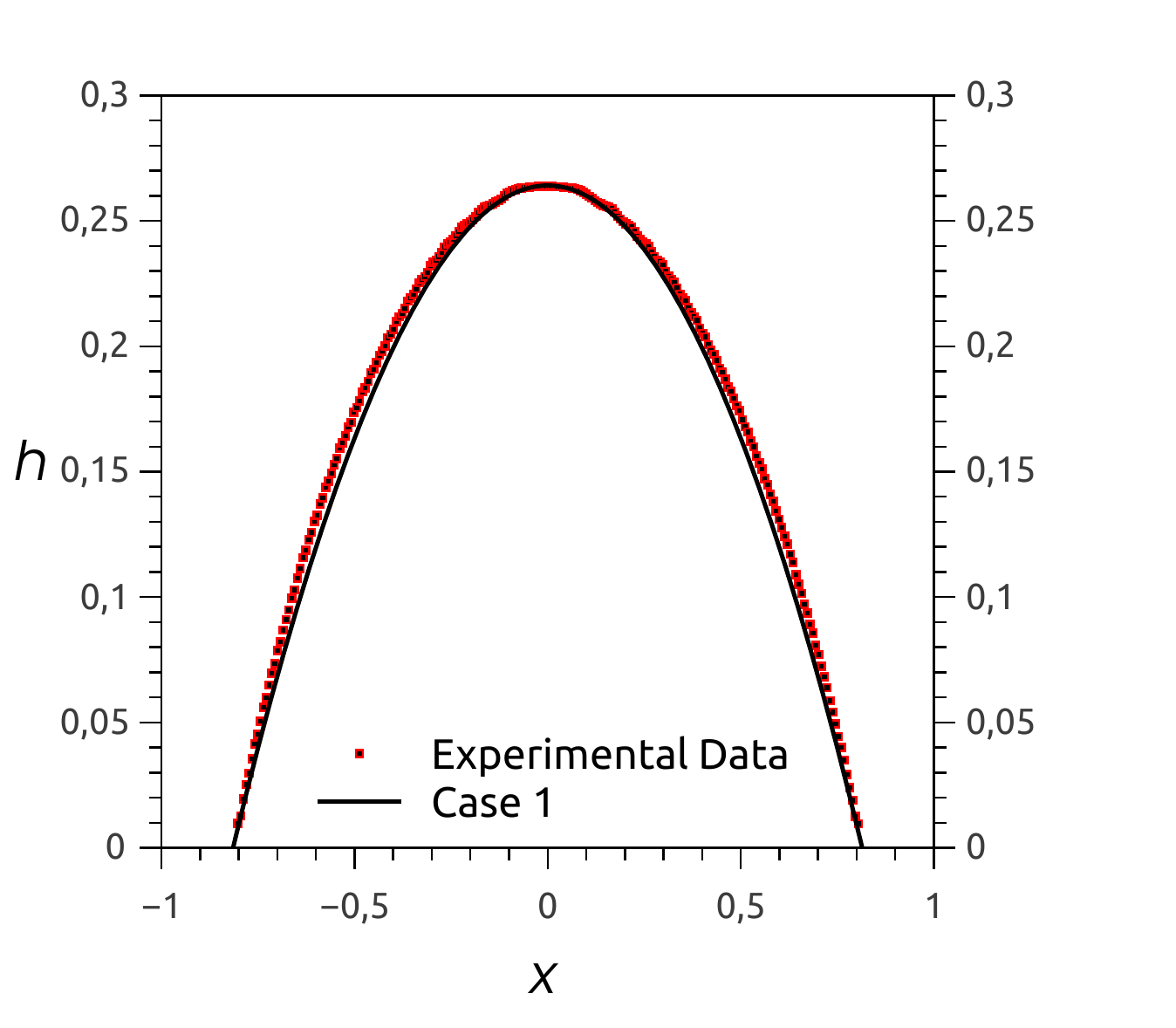}
\label{heightPolarC1}}
\subfigure []
{\includegraphics[width=0.3\linewidth]{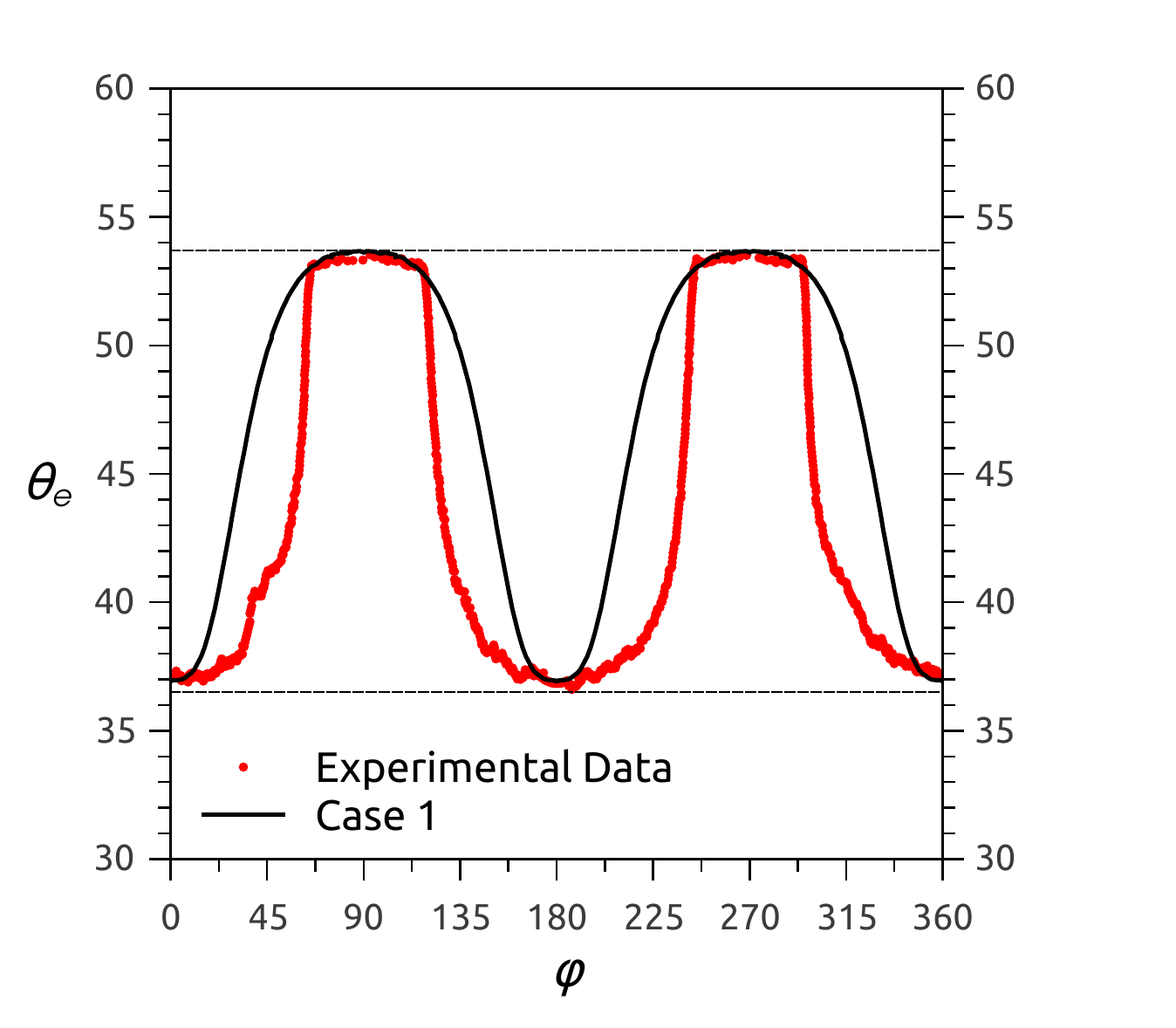}
\label{AngPolarC1}}
\caption{Comparison between the experimental data (red points) and the polar solution (black dashed--dotted lines) determined with the \emph {four parameters of Case $1$} for: (a) drop footprint, (b) longitudinal thickness profile along the filament ($x$--axis), and (c) contact angle distribution around the footprint periphery. The horizontal dashed lines correspond to $\theta_{rcd}$ and $\theta_{adv}$ shown in Fig.~\ref{fig:hystCycle}.}
\label{fig:Exp_Case1}
\end{figure}

Since it is clear now that a truncated polar solution is  appropriate to fit the experimental data except for some details, we analyze how to select a minimal and optimal set of measured parameters necessary to have a good description of the drop. This can be useful when the experimental setup does not allow a detailed measurement of five of them given in Eq.~(\ref{eq:condic}). With this goal, we consider the following different combinations of four parameters:

{\it Case 1}: Measure $(w_x,w_y,h_{max},\theta_x)$, and take $\theta_y$ equal to the advancing contact angle obtained in the hysteresis cycle (Fig. \ref{fig:hystCycle}) over dry substrate, i.e. $\theta_y=\theta_{adv}=53.7^\circ$ instead of the measured value of $\theta_y$ for the drop itself. The comparison with the experimental data is shown in Fig.~\ref{fig:Exp_Case1}. Clearly, the results for the footprint, the thickness profile, and the azimuthal distribution of $\theta_e$  remain with an  accuracy very similar to that in Fig.~\ref{fig:Exp_Polar} for five measured parameters. The only minor difference is in the largest contact angle, which is due to the fact that $\theta_y$ taken from the hysteresis cycle is $0.15^\circ$ larger than the one measured in this particular case.

\begin{figure}[ht]
\subfigure []
{\includegraphics[width=0.3\linewidth]{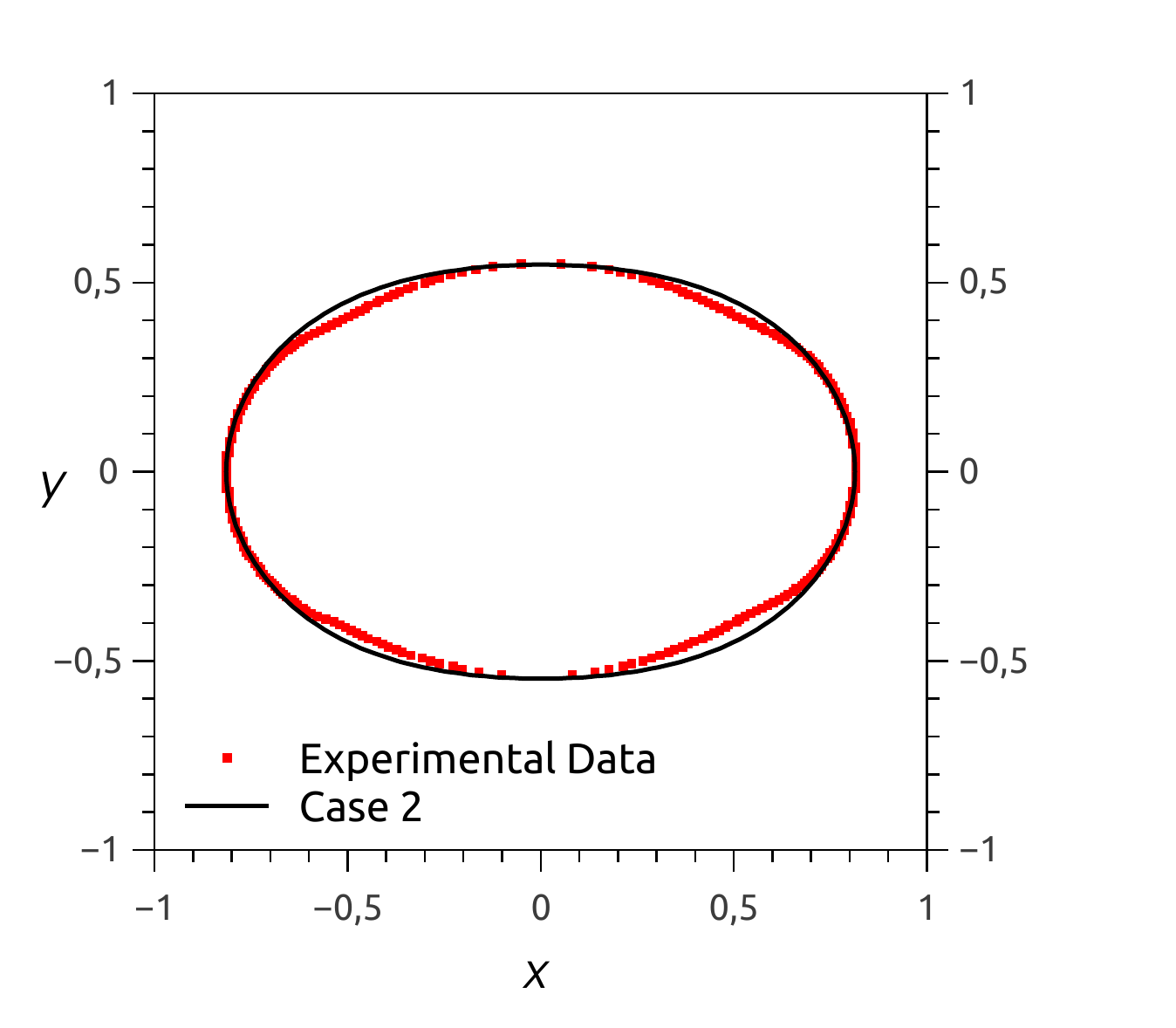}
\label{FootPolarC2}}
\subfigure []
{\includegraphics[width=0.3\linewidth]{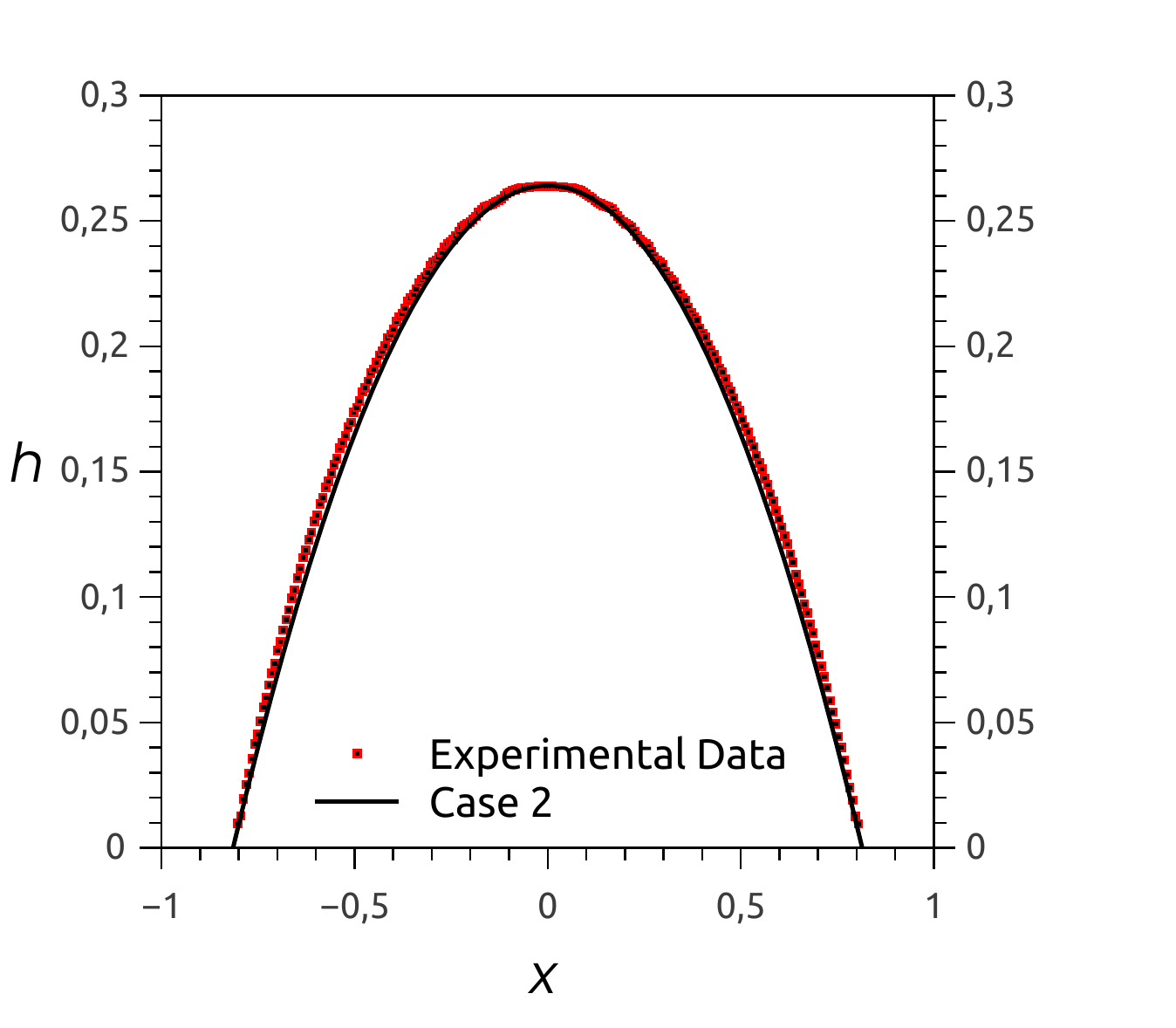}
\label{heightPolarC2}}
\subfigure []
{\includegraphics[width=0.3\linewidth]{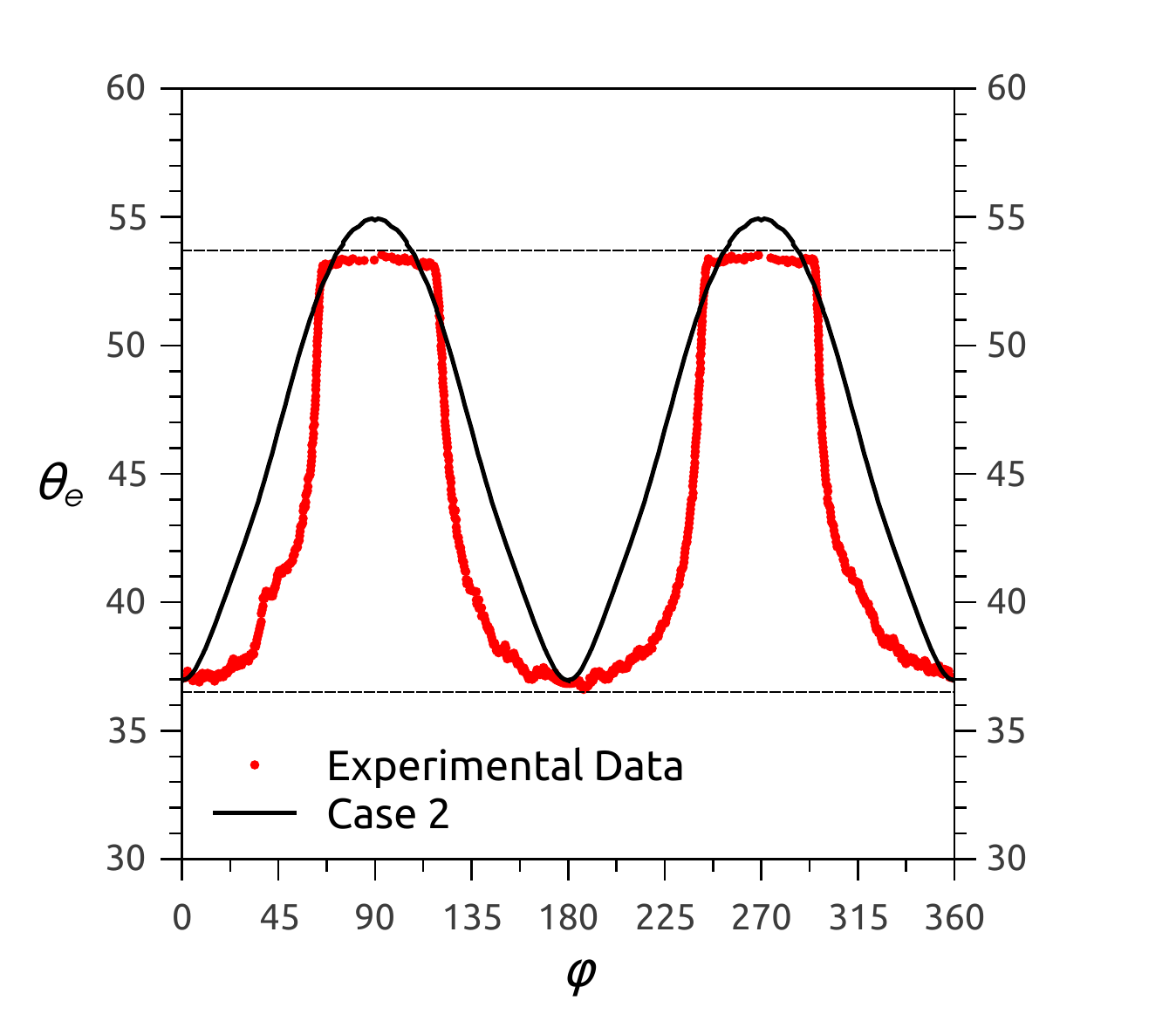}
\label{AngPolarC2}}
\caption{Comparison between the experimental data (red points) and the polar solution (black dashed--dotted lines) determined with the \emph {four parameters of Case $2$} for: (a) drop footprint, (b) longitudinal thickness profile along the filament ($x$--axis), and (c) contact angle distribution around the footprint periphery. The horizontal dashed lines correspond to $\theta_{rcd}$ and $\theta_{adv}$ shown in Fig.~\ref{fig:hystCycle}.}
\label{fig:Exp_Case2}
\end{figure}

{\it Case 2}: Measure $(w_x,w_y,h_{max},\theta_x)$, and calculate $\theta_y$ from Eq.~(\ref{eq:thetavsw}). Figure~\ref{fig:Exp_Case2} shows the comparison with experiments. While the thickness profile in Fig.~\ref{heightPolarC2} does not show any significant departure from the experimental data, the comparison of the footprint shape in Fig.~\ref{FootPolarC2} is worse than before. Similarly to Case $1$, the angular distribution of the contact angle in Fig.~\ref{AngPolarC2} shows larger differences. In this case, the use of Eq.~(\ref{eq:thetavsw}) overestimates $\theta_y$ in about $2.6\%$ respect to the measured value. This difference, though seemingly small, is important since it prevents the solution from giving a plateau region at the maximum contact angle. Also, it affects the curvature of the transition region towards the minimum.

\begin{figure}[ht]
\subfigure []
{\includegraphics[width=0.3\linewidth]{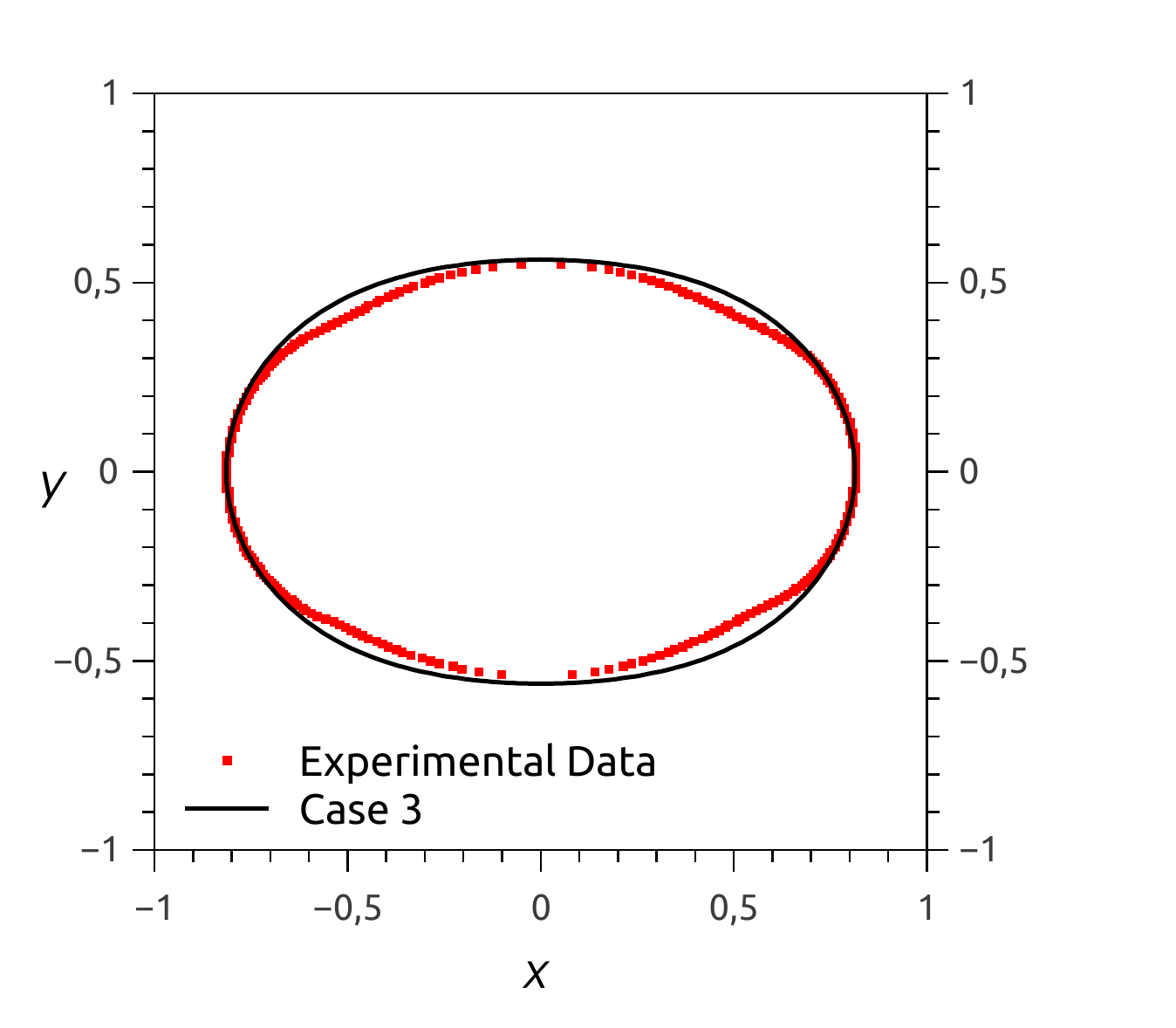}
\label{FootPolarC3}}
\subfigure []
{\includegraphics[width=0.3\linewidth]{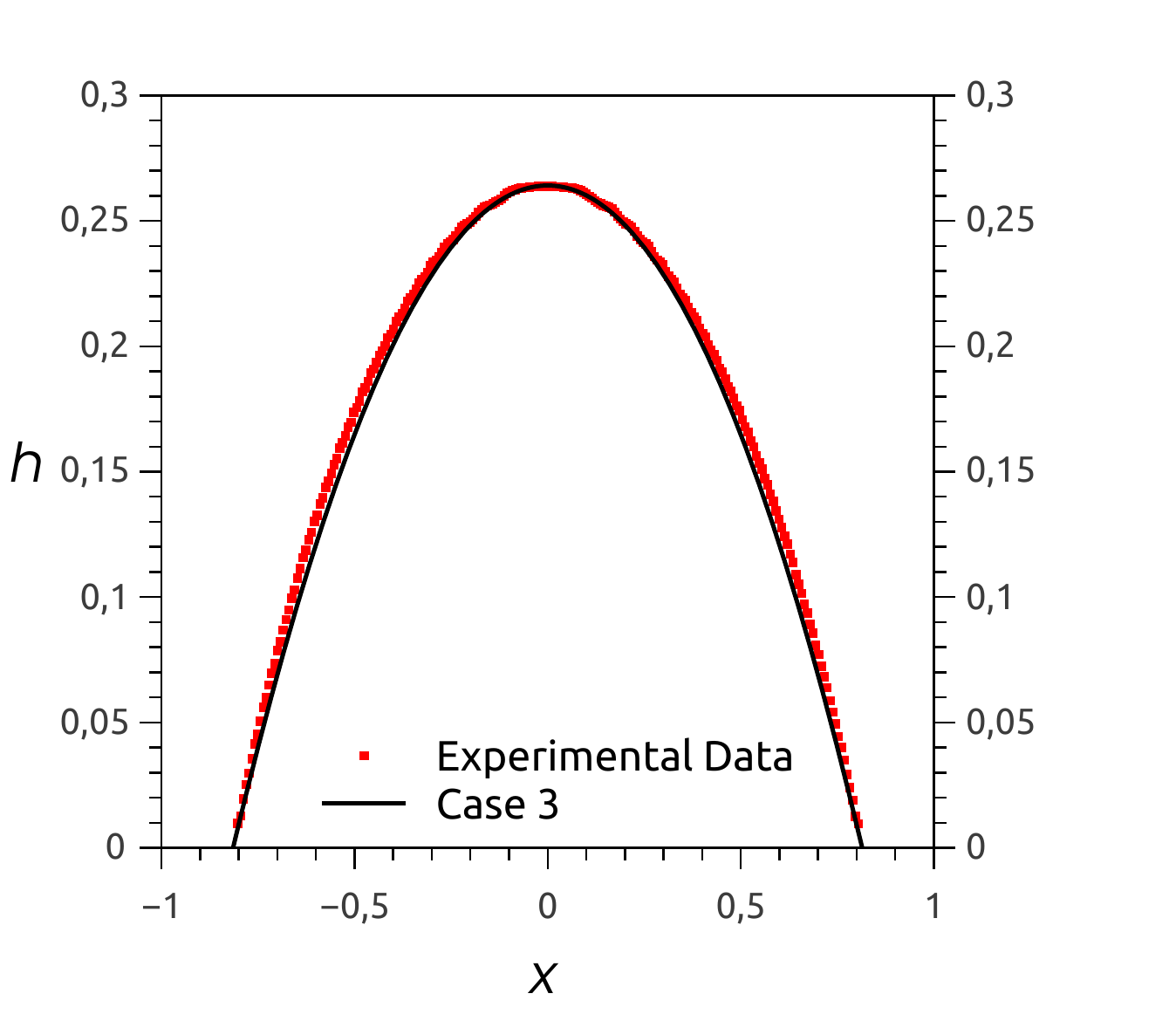}
\label{heightPolarC3}}
\subfigure []
{\includegraphics[width=0.3\linewidth]{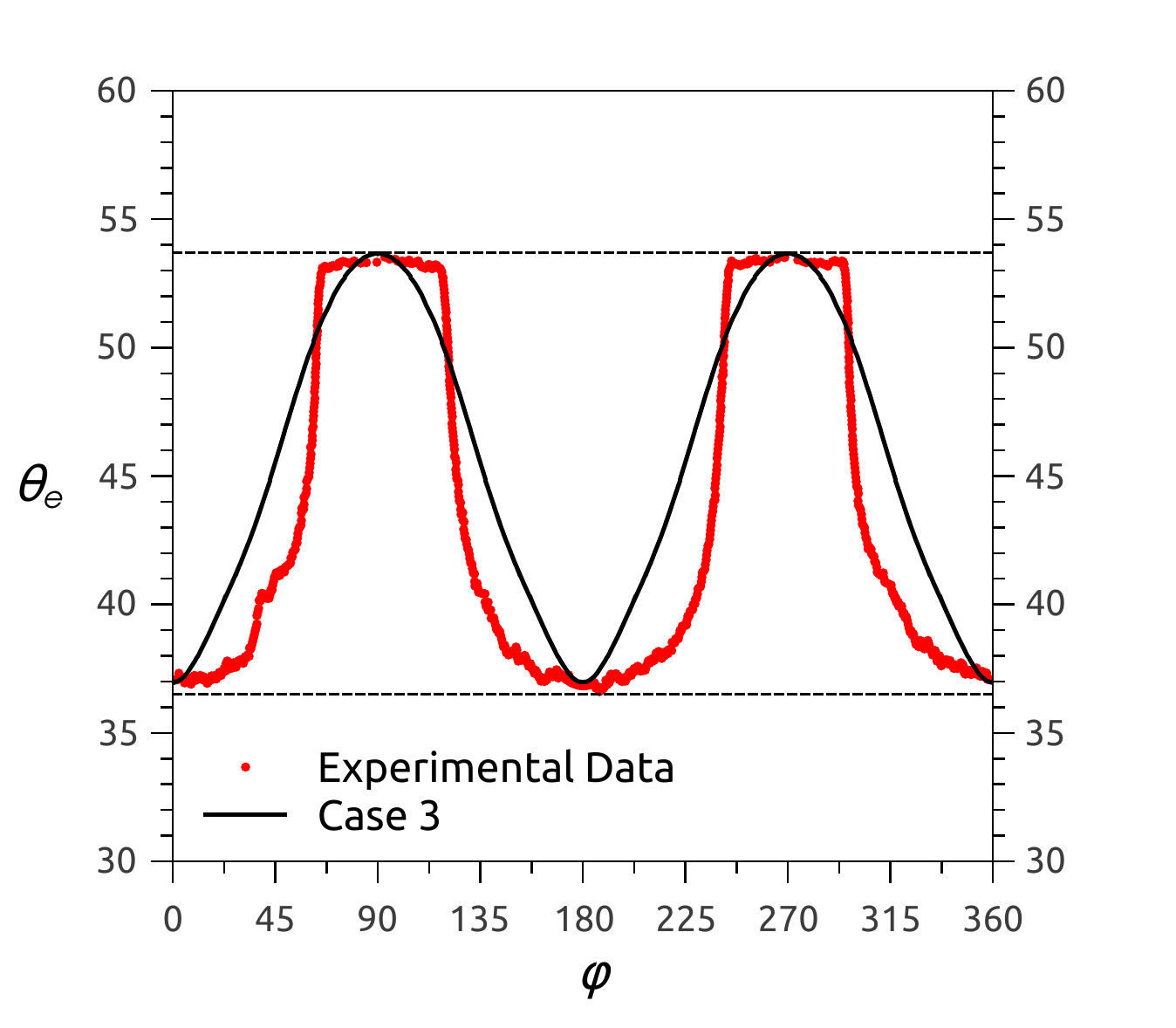}
\label{AngPolarC3}}
\caption{Comparison between the experimental data (red points) and the polar solution (black dashed--dotted lines) determined with the \emph {four parameters of Case $3$} for: (a) drop footprint, (b) longitudinal thickness profile along the filament ($x$--axis), and (c) contact angle distribution around the footprint periphery. The horizontal dashed lines correspond to $\theta_{rcd}$ and $\theta_{adv}$ shown in Fig.~\ref{fig:hystCycle}.}
\label{fig:Exp_Case3}
\end{figure}

{\it Case 3}: Measure $(w_x,h_{max},\theta_x,\theta_y)$, and calculate $w_y$ from Eq.~(\ref{eq:thetavsw}) (see Fig.~\ref{fig:Exp_Case3}). The footprint shown in Fig.~\ref{FootPolarC3} is a bit worse than that for Case $2$ basically because now there is on overestimation of about $1.9\%$ for $w_y$ from the use of Eq.~(\ref{eq:thetavsw}). The thickness profile in Fig.~\ref{heightPolarC3} is again in very good agreement with the experimental data, and the angular distribution of the contact angle in Fig.~\ref{AngPolarC3} is better than in Case $2$, since no over shooting is obtained at the maximum region. Note that if $\theta_y$ were taken from  the hysteresis cycle as being equal $\theta_{adv}$, instead of using Eq.~(\ref{eq:thetavsw}), then these results could have been obtained without the need of the top--view picture, since the information in the side--view picture is enough. Thus, in this case, the number of measured values for the polar solution can be reduced to only three, as it happens in the Cartesian one.

\begin{figure}[ht]
\subfigure []
{\includegraphics[width=0.3\linewidth]{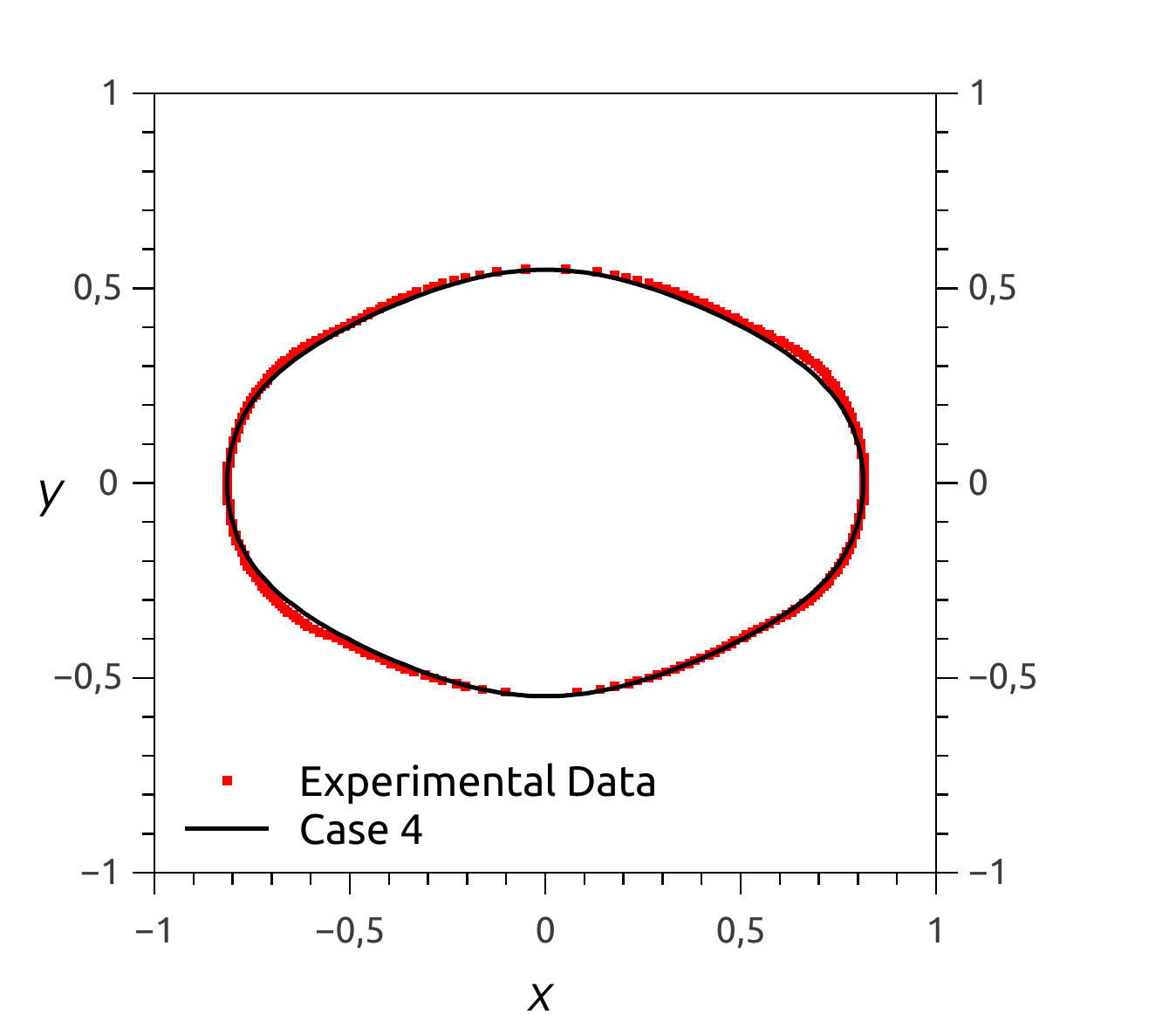}
\label{FootPolarC4}}
\subfigure []
{\includegraphics[width=0.3\linewidth]{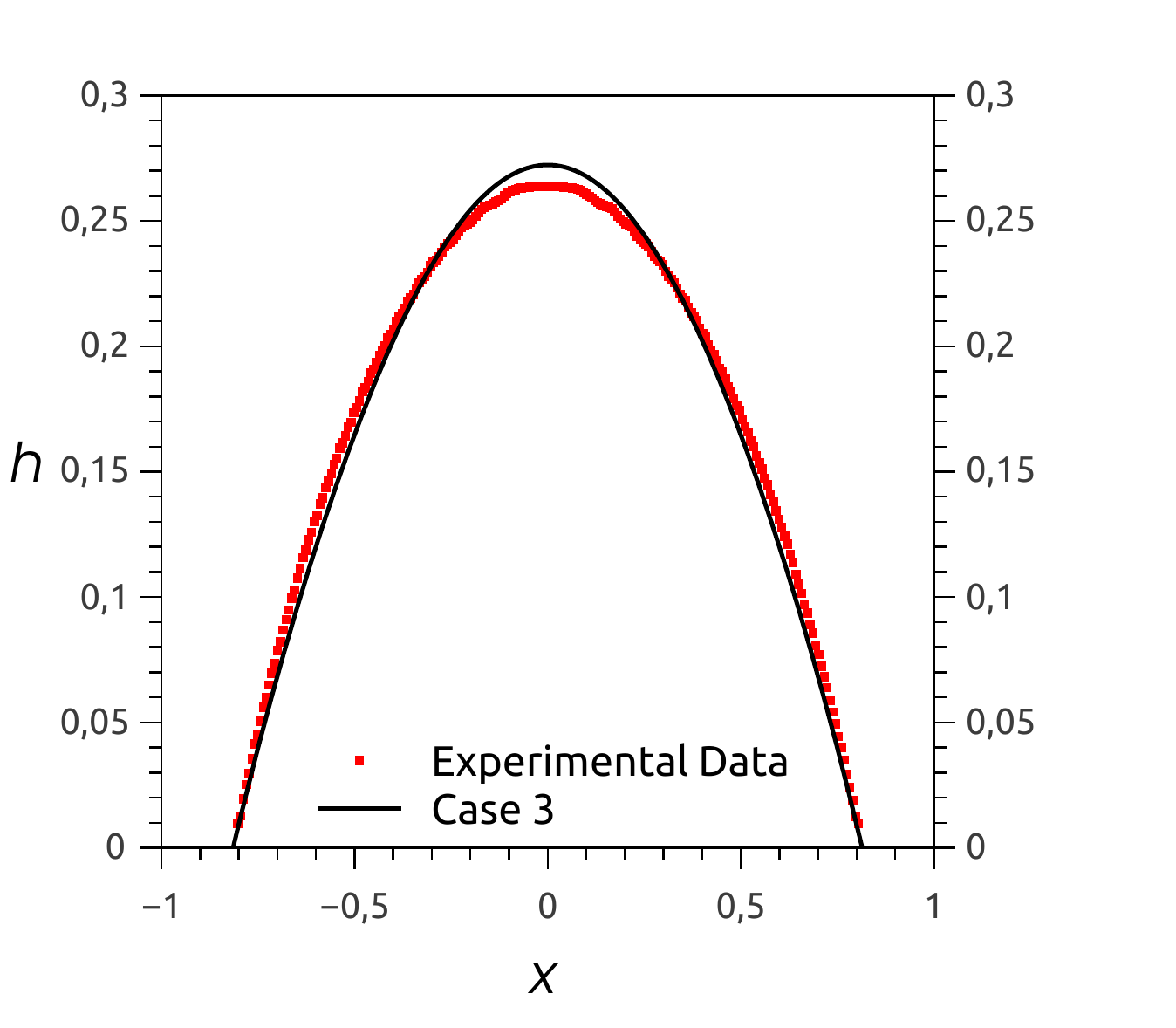}
\label{heightPolarC4}}
\subfigure []
{\includegraphics[width=0.3\linewidth]{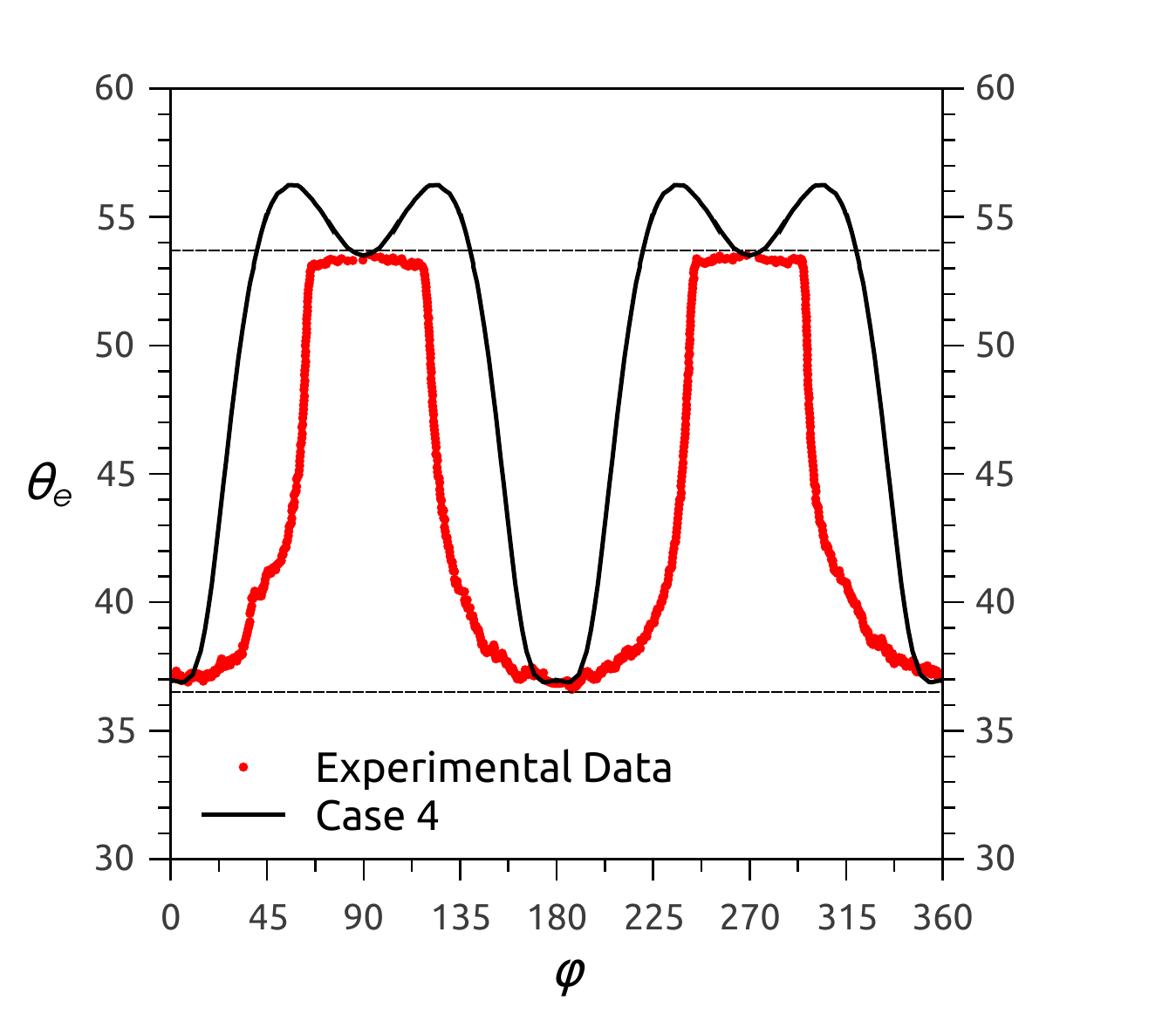}
\label{AngPolarC4}}
\caption{Comparison between the experimental data (red points) and the polar solution (black dashed--dotted lines) determined with the \emph {four parameters of Case $4$} for: (a) drop footprint, (b) longitudinal thickness profile along the filament ($x$--axis), and (c) contact angle distribution around the footprint periphery. The horizontal dashed lines correspond to $\theta_{rcd}$ and $\theta_{adv}$ shown in Fig.~\ref{fig:hystCycle}.}
\label{fig:Exp_Case4}
\end{figure}

{\it Case 4}: Measure $(w_x,w_y,\theta_x,\theta_y)$, and calculate $h_{max}$ by approximating the thickness profile along $x$--axis with an arc of a circle (see Fig.~\ref{fig:Exp_Case4}). In fact, from the values of $w_x$ and $\theta_x$ it is possible to approximate the longitudinal thickness profile with an arc of a circle, and then the value of $h_{max}$ can be estimated with an error of the order of $3\%$ respect to the actually measured value. The quality of the adjustment can be seen in Fig.~\ref{heightPolarC4}. The adjustment of the footprint is very good because all parameters used for its calculation are measured. However, the small difference in $h_{max}$ has strong consequences in the angular distribution of $\theta_e$ (see Fig. \ref{AngPolarC4}). Even if the solution is in agreement with the experimental data in the region around the minimum, the transition region presents an unexpected behavior despite the fact that $\theta_y$ is measured. The most remarkable difference is in the zone of the maximum, where two peaks with larger contact angles show up instead of the expected plateau. This result suggests that the calculated solution for this case does not lead to a smooth surface, and warns us about the importance of using an accurate value for $h_{max}$ for the prediction of the azimuthal distribution of the contact angle.

\section{Full numerical simulations}
\label{sec:num}

Since both analytical solutions, the Cartesian and the polar ones, were obtained within the framework of the lubrication approximation, which is valid for small free surface slopes and contact angles, we perform in this section numerical simulations of the full Navier--Stokes equation. 

In order to obtain the shape of the sessile drop, we calculate the evolution of a liquid filament on a partially wetting substrate by solving numerically Eq.~(\ref{eq:NS}). We simulate the real experimental configuration that actually leads to the drop under study without any assumption about the size of the contact angles. Several particular characteristics of the flow must be taken into account to perform this simulation. Firstly, the well known divergence at moving contact lines must be overcome. This is done by relaxing the no slip boundary condition at the substrate through the Navier formulation,
\begin{equation}
 v_{x,y} = \ell \, \frac{\partial v_{x,y}}{\partial z},
\end{equation}
where $\ell$ is the slip length. Secondly, the filament must be short enough to avoid pinching necks along it, because this change of topology is not supported by regular numerical schemes~\cite{ubal_pof2014}. Thus, the filament must evolve into a single drop, and its volume must then be equal to that of the resulting drop. Here, we consider that the initial filament can be described by a cylindrical cap of length $L$ and width $w$, ended by two spherical caps. The curvature radius, $R$, of both the cylinder and the spheres is given by, $R=w/(2\sin \theta_0)$, where $\theta_0$ is initial contact angle of the filament. Consequently, the thickness of the filament is $h_0=R (1-\cos \theta_0)$.

Finally, the hysteretic behavior of the contact angle is introduced by ascribing different values of $\theta_e$ to different regions of the substrate. Thus, we consider that the region originally occupied by the filament (the rectangular area limited by $-w \leq x \leq L+ w$, and $-w/2 \leq y \leq w/2$) has $\theta_e=\theta_{rcd}$, while for the rest of the substrate it corresponds $\theta_e=\theta_{adv}$. This spatial distribution of $\theta_e$ assumes that the contact line will only recede in the region where the filament initially sits, and only advances outside of it. Naturally, the step function that this description implies should be slightly smoothed to avoid numerical issues. Moreover, the experiments show that in a region close and outside of the filament contact line both kinds of motion may take place. Therefore, we also consider a wider smooth transition of $\theta_e$ along $y$-direction (normal to the filament) within a certain distance $\delta$, which should be of the order of $w$. Thus, we construct the wettability of the substrate as:
\begin{equation}
\theta_e(x,y)= \theta_x(x) \, \theta_y(y),
\end{equation}
where
\begin{eqnarray}
 \theta_x &=& \frac{\tanh \left[ q_x (x+w/2) \right]- \tanh \left[ q_x (x-L-w/2) \right]}{2} \nonumber \\
 \theta_y &=& \frac{\tanh \left[ q_y (x+w/2+\delta/2) \right]- \tanh \left[ q_y (y-w/2-\delta/2) \right]}{2},
\end{eqnarray}
where $q_x$ and $q_y$ set the average slope of the transition region. Here, we take $q_x=100$, $q_y=50$, and $\delta =1.1 w/2$.

We use a Finite Element technique in a domain which deforms with the moving fluid interface by using the Arbitrary Lagrangian-Eulerian (ALE) formulation. The interface displacement is smoothly propagated throughout the domain mesh using the Winslow smoothing algorithm. The main advantage of this technique compared to others such as the Level Set of Phase Field techniques is that the fluid interface is and remains sharp. The main drawback, on the other hand, is that the mesh connectivity must remain the same, which precludes the modeling of situations for which the topology might change (rupture of the filament). The default mesh used throughout is unstructured, and typically has $3\times 10^4$ triangular elements (linear elements for both velocity and pressure). Automatic remeshing is enabled to allow the solution to proceed even for large domain deformation when the mesh becomes severely distorted. The mesh nodes are constrained to the plane of the boundary they belong to for all but the free surface.

\begin{figure}
\centering
\subfigure[]
{\includegraphics[width=0.8\textwidth]{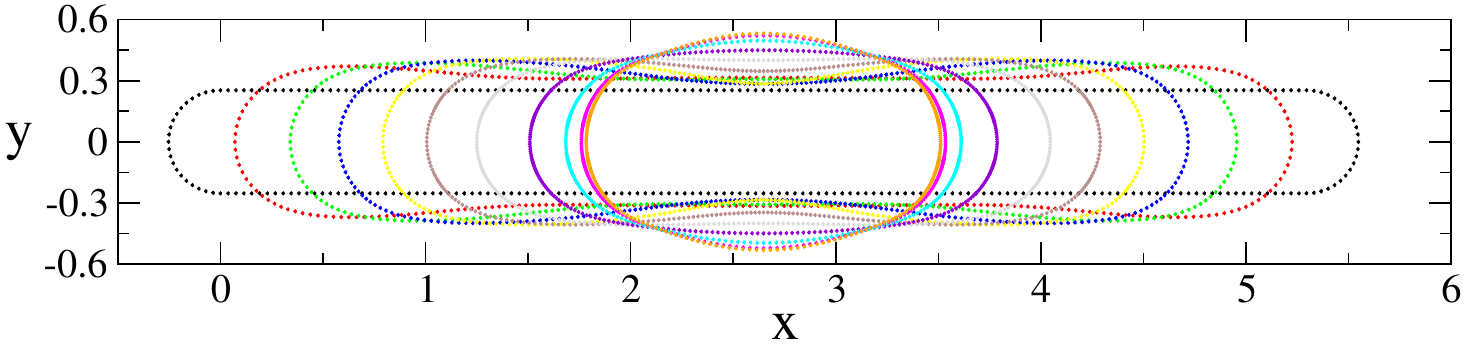}\label{fig:foots}}
\subfigure[]
{\includegraphics[width=0.8\textwidth]{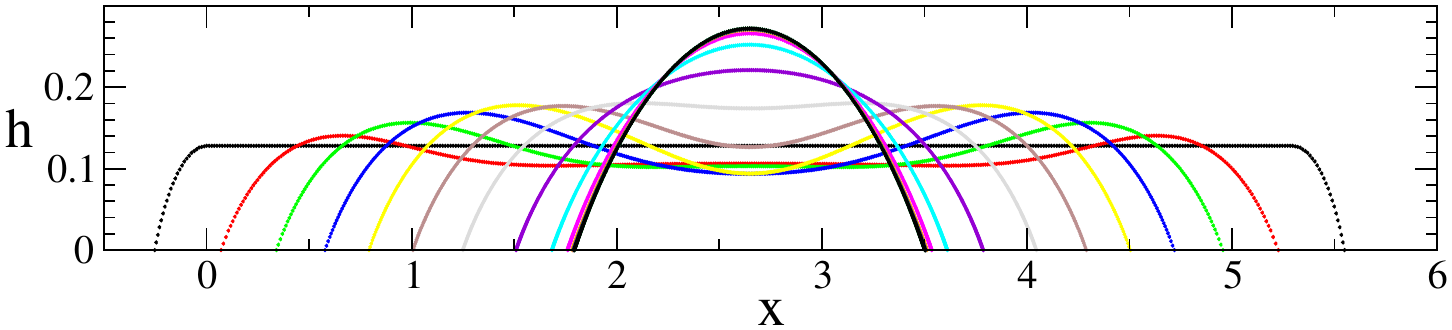}\label{fig:hx_t0-200}}
\caption{Time evolution of: (a) the contact line (footprint), and (b) the thickness profile along the filament axis ($y=0$), from $t=0$ up to $t=200$ for a time step $\Delta t=20$.}
\label{fig:foots-hx}
\end{figure}

To simulate the evolution of a short filament that leads to a single drop, we calculate the filament length, $L$, for a given experimental value of the width, $w$. This can be easily done by knowing the drop volume, $V_{drop}$, under study. For the case in Fig.~\ref{fig:evolFilam}, we have $w=0.5062$ and, with $\theta_0=57.3^\circ$, we find $L=5.296$. Assuming $\theta_{rcd}=36.5^\circ$ and $\theta_{adv}=53.7^\circ$, we obtain the time evolution of the footprint shown in Fig.~\ref{fig:foots}. In these simulations we use $La=0.005$, as given by the experimental parameters, and set $\ell=0.005$. The runs with smaller values of $\ell$ produced only negligible differences.

At early times, the ends of the filament dewet and a bulge is formed at each end. So, the side parts of the buldge contact line lie outside the filament footprint: the most external portions of this contact line dewet, while the most internal ones advance towards the center of the filament ($x=L/2$). This is in agreement with experimental observations, and justifies the choice of a wide transition region for $\theta_y$, which yields intermediate angles in the region close and outside the filament footprint. 

In Fig.~\ref{fig:hx_t0-200}, we plot the thickness profile along the meridian plane ($y=0$). At early times, a depression appears at the middle of the filament, but since $L$ is relatively short, the behavior of the thickness there changes, and it starts growing for later times, thus building up the central region of the drop.

For very large times, say $t=200$, the drop reaches a static shape at rest, and adopts a footprint very close to that in the experiments. We compare in the following section the numerical results thus obtained with the theoretical solutions within the lubrication approximation as well as the experiments.

\begin{figure}
\subfigure []
{\includegraphics[width=0.3\linewidth]{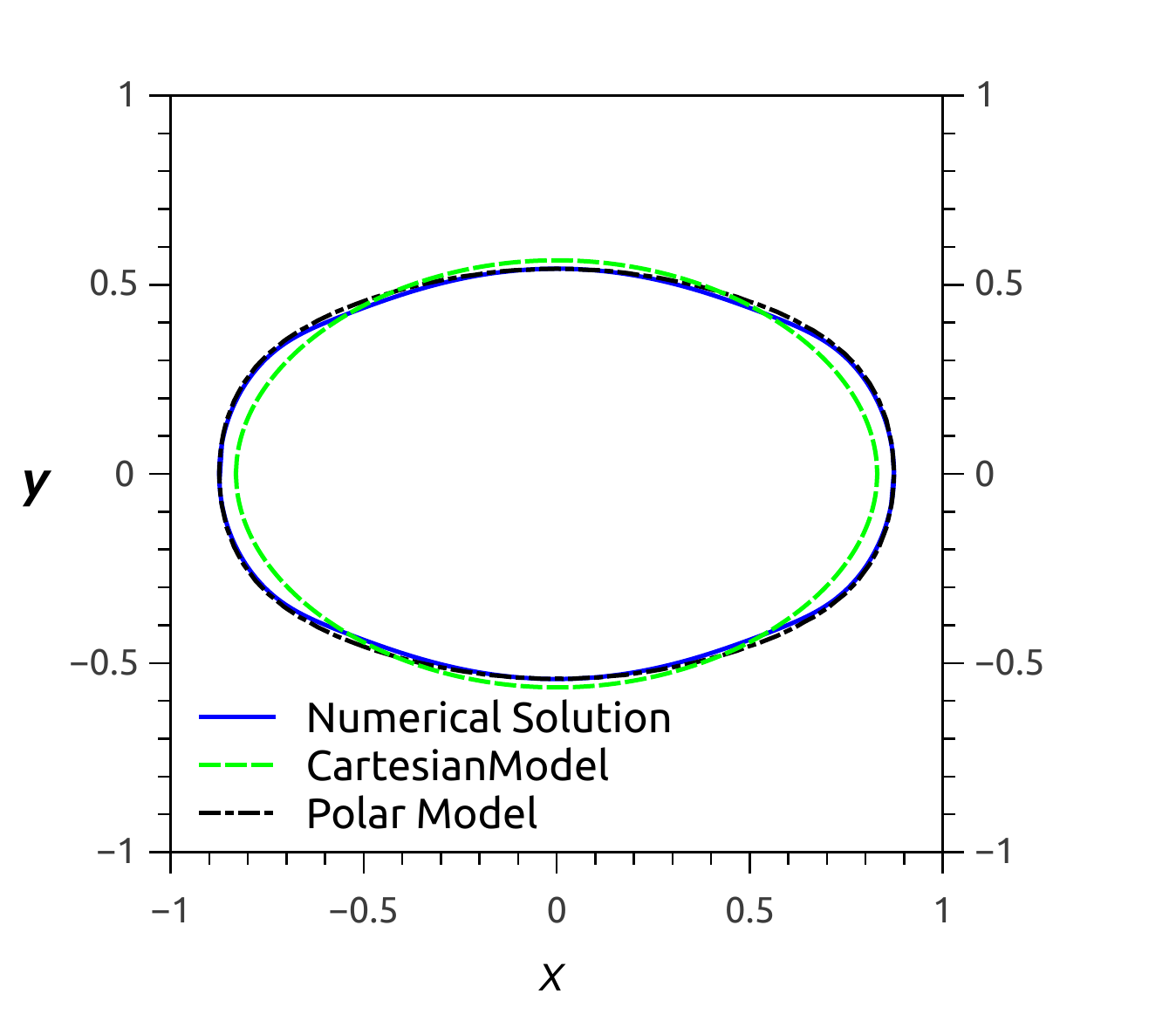}\label{FootGotaTeor}}
\subfigure []
{\includegraphics[width=0.3\linewidth]{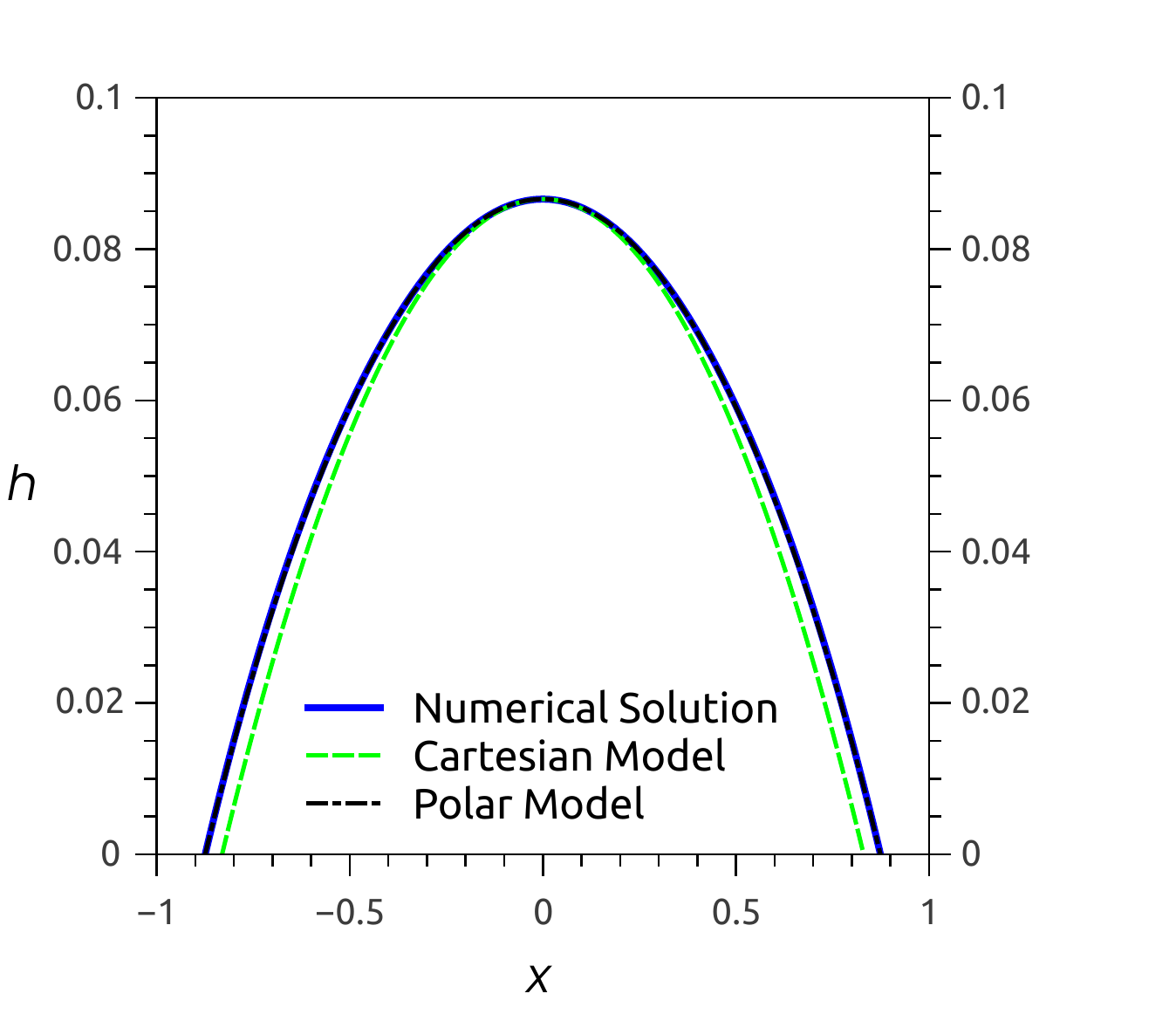}\label{hxGotaTeor}}
\subfigure []
{\includegraphics[width=0.3\linewidth]{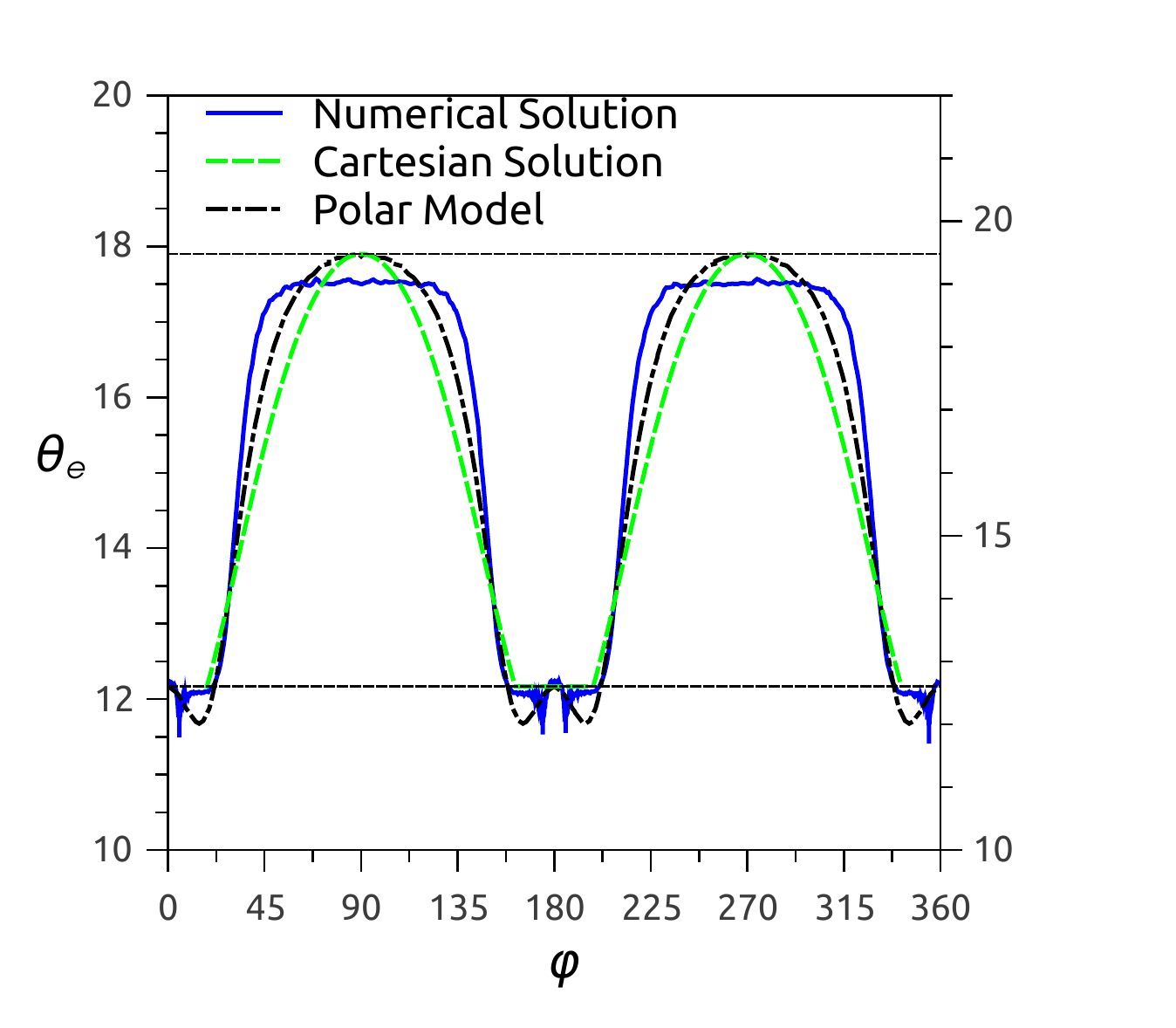}\label{AnglGotaTeor}}
\caption{Comparison between the numerical footprint (dot-dashed blue line) with the Cartesian (dashed green line) and polar (black dot-dashed line) solutions for \emph {small contact angles}: (a) drop footprint, (b) longitudinal height profile along the filament ($x$--axis), and (c) contact angle distribution around the footprint periphery. The horizontal dashed lines correspond to $\theta_{rcd}/3$ and $\theta_{adv}/3$ shown in Fig.~\ref{fig:hystCycle}.}
\label{fig:small_ang}
\end{figure}

In order to validate the numerical simulations, we first compare its results with those of the lubrication approximation solutions. To do so, we perform calculations with smaller contact angles as those in the experiments, namely one third of the values defined above, i.e. we use here: $\theta_{adv}=53.7^\circ/3=17.9^\circ$ and $\theta_{rcd}=36.5^\circ/3=12.16^\circ$. The comparison is shown in Fig.~\ref{fig:small_ang} (for the Cartesian solution we use $\theta_x$, $\theta_y$, and $h_{max}$ as input parameters). As expected, very little differences exist among the three curves for the footprints in Fig.~\ref{FootGotaTeor}, and the longitudinal thickness profiles in  Fig.~\ref{hxGotaTeor}. Interestingly, the numerical simulations show in Fig.~\ref{AnglGotaTeor} a steeper transition between the maximum and minimum contact angles than those given by the long wave solutions (the small difference of $1^\circ$ at the maximum of the numerical is probably related to a grid effect). Therefore, except for these details in the angular distribution, the simulations as posed can be safely used to describe the physical problem in hand.

\begin{figure}
\subfigure []
{\includegraphics[width=0.3\linewidth]{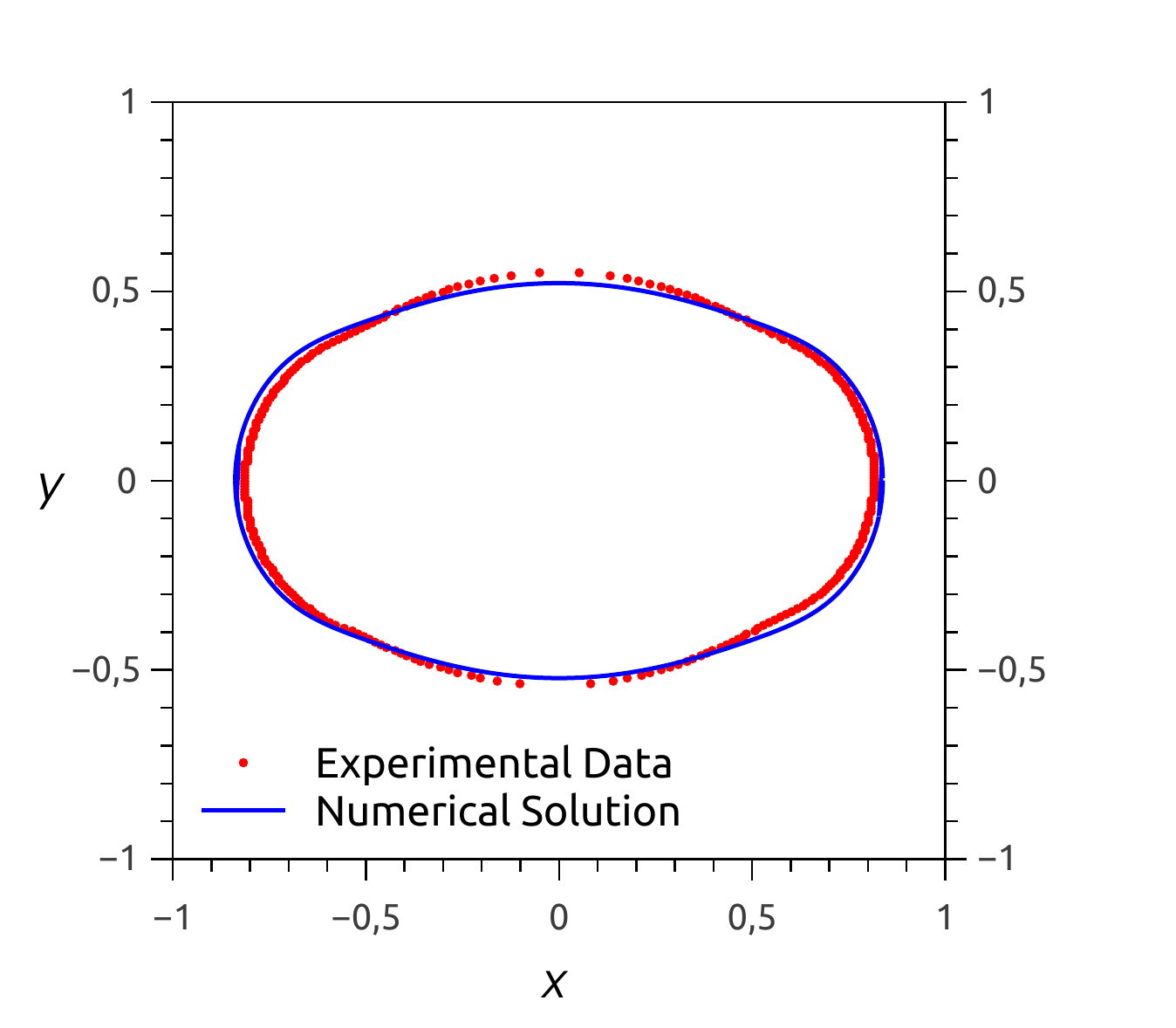}\label{FootGotaEXPyNUM}}
\subfigure []
{\includegraphics[width=0.3\linewidth]{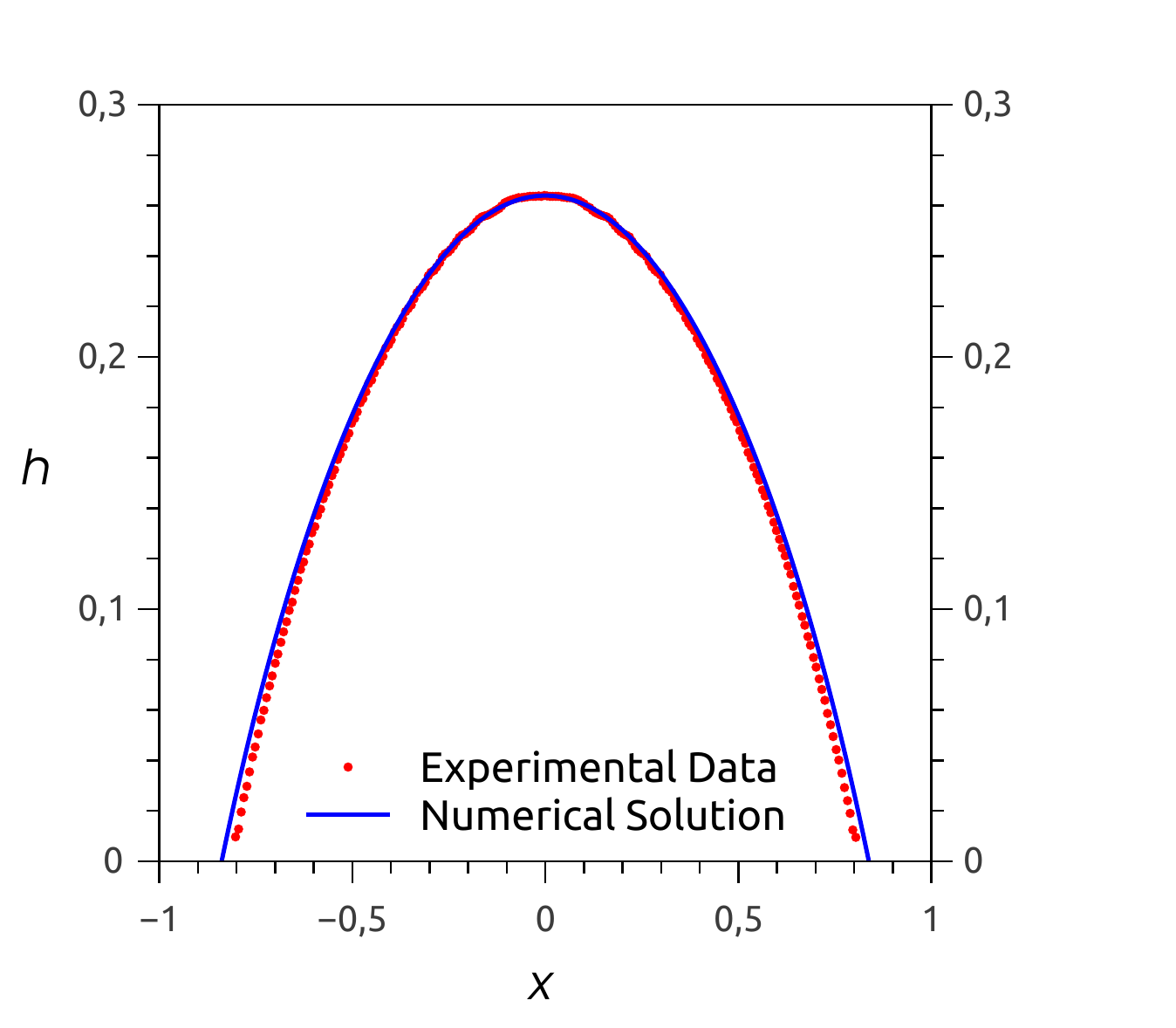}\label{hxGotaEXPyNUM}}
\subfigure []
{\includegraphics[width=0.3\linewidth]{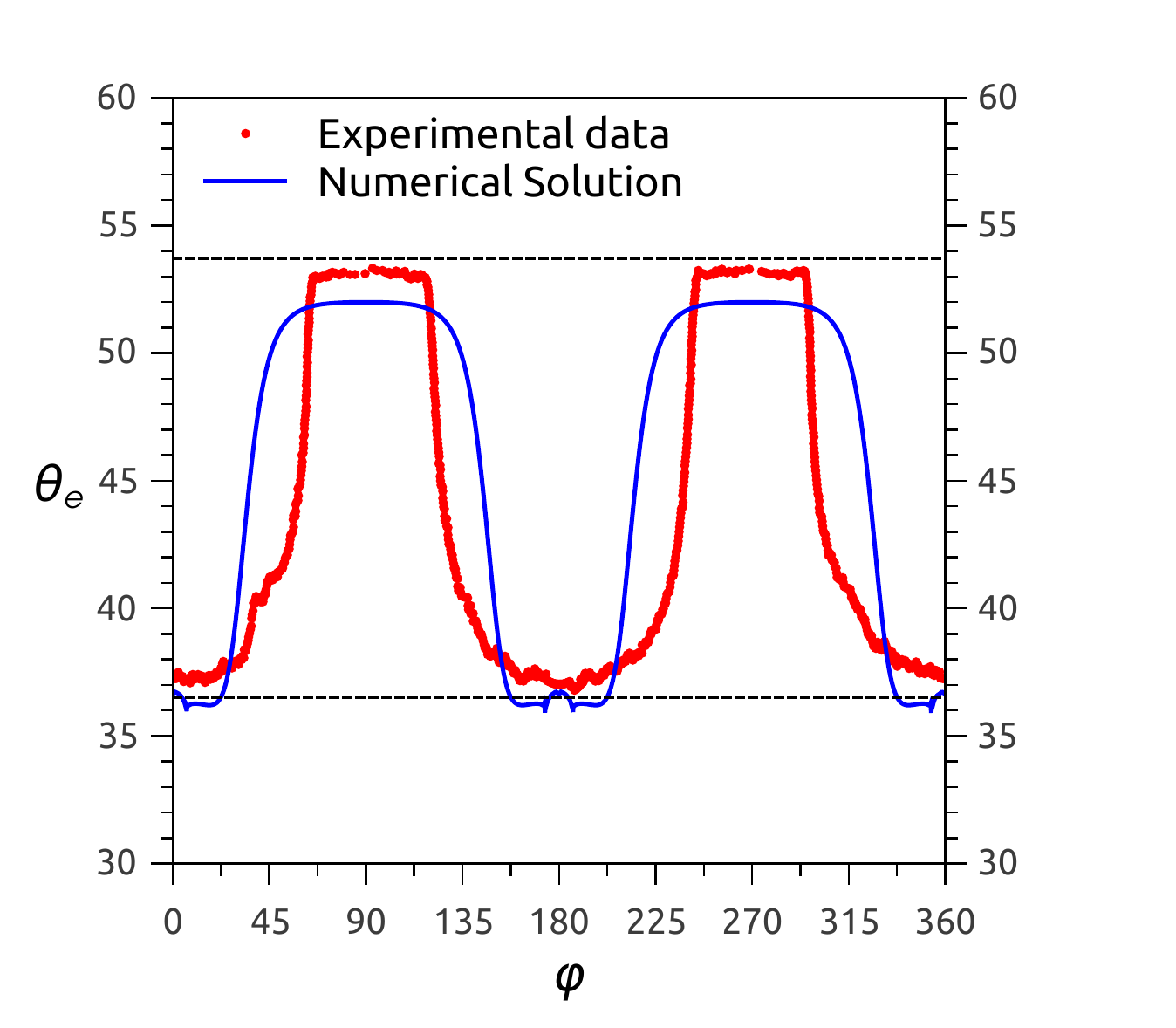}\label{AnglGotaEXPyNUM}}
\caption{Comparison between the experimental data (red points) and the numerical simulations (solid blue lines) for: (a) drop footprint, (b) longitudinal thickness profile along the filament ($x$--axis), and (c) contact angle distribution around the footprint periphery. The horizontal dashed lines correspond to $\theta_{rcd}$ and $\theta_{adv}$ shown in Fig.~\ref{fig:hystCycle}.}
\label{fig:num-exp}
\end{figure}

Now,  we compare in Fig.~\ref{fig:num-exp} the numerical solution for large times with the experimental data for the drop with large contact angles. The footprints comparison in Fig.~\ref{FootGotaEXPyNUM} shows that the simulations are able to account for the slight change of curvature sign in the regions connecting the receding and advancing apexes, which is a characteristic feature of this type of drops. However, the simulation   overestimates the value of the $w_x$ in about $3.7\%$, and consequently, underestimates $w_y$ in the same amount. However, this small difference in $w_x$ does not have an important effect on the comparison with the thickness profile, as seen in Fig.~\ref{hxGotaEXPyNUM}).  

As regards to the dependence $\theta_e(\varphi)$,  Fig.~\ref{AnglGotaEXPyNUM} shows that the simulation yields an angular distribution similar to that of the experiments in the sense that both show steep transitions between the maximum ($\theta_e \approx \theta_{adv}$) and minimum ($\theta_e \approx \theta_{rcd}$), and a relatively wider plateau at the maximum region. The differences between numerics and experiments at these extremes ($\sim 1^\circ$) are due to numerical errors. Interestingly, the equality of ratios in Eq.~(\ref{eq:thetavsw}) is satisfied within an error of $1.5\%$, which means that the departure of widths and angles are consistent with this basic property of the solution. In general, this comparison shows that the form used here to emulate the hysterical behavior of the contact angle is appropriate to study this kind of problems.

\label{sec:exp1}

\section{Summary and conclusions}
\label{sec:conc}

In this work we study in detail the morphology of sessile drops with non--circular footprints. In particular, we focus on the kind of drops that result from the breakup of a liquid filament placed on a horizontal substrate under partial wetting conditions. In order to obtain this type of wetting when the liquid (silicon oil PDMS) is in contact with the glass substrate, we modify its surface energy by coating it with a fluorinated solution. The resulting wettability is characterized by a hysteresis curve for the contact angle, so that the advancing and receding angles are known parameters of the system while modeling of the flow. By means of optical techniques, we are able to measure the drop footprint, the thickness profile along the filament axis, and the azimuthal distribution of contact angle around the drop periphery.

The shape of a typical static droplet thus obtained is described by using the long wave theory, or lubrication approximation. Two analytical solutions,  in Cartesian and polar coordinates, are found within this framework. The main goal of looking for these solutions is to be able to predict details of the drop that are difficult to measure, such as the shape of the footprint, the thickness profiles, and the angular distribution of contact angles. We determine these solutions by using a minumum number of easily measured parameters, such as maximum or minimum diameters, angles and thicknesses. The solutions thus obtained are contrasted with detailed experimental measurements to assess their accuracy.

While the Cartesian solution is restricted by the choice of at most three (measurable) parameters, the polar one is truncated to allow for five of them. In this sense, we also study four different ways to combine the parameters necessary to fix the polar solution. Interestingly, the Cartesian solution provides a simple relationship between the widths of the drop and the contact angles along the axes of symmetry, which holds with good approximation when compared with experimental data. In general, the good agreement found in the comparison with the experiments confirms that the lubrication approximation solutions are quantitatively suitable to describe the drops with non--circular footprints.

Since the experimental drops have relatively large contact angles, and the theoretical modeling assumes small values of the free surface slopes, there might remain some doubts about the comparison with these experiments. So, in order to put in evidence possible effects related to the presence of large contact angles, we also compute the drop shape by performing time evolving simulations of the full Navier--Stokes equation. In these calculations, two main features have to be included: a) the relaxation of the contact line singularity by including a slip model at the boundary condition on the substrate, and b) the hysteresis of the contact angle. The latter is emulated by imposing a specific spatial distribution of contact angle as suggested by experimental observations, that is, by setting $\theta_{rcd}$ in the region initially occupied by the filament, and $\theta_{adv}$ outside of it, with a smoothed and narrow transition region. The simulations are validated with the theoretical solutions for small contact angles, and are also compared with the experimental data. The most remarkable result of these simulations is that the effect of having large contact angles is to widen the regions of constant $\theta_e$ around the minimum ($\theta_e \approx \theta_{rcd}$) and maximum ($\theta_e \approx \theta_{adv}$) values, as well as to make steeper the transitions regions in this angular diagram. A similar effect is also observed with the polar solution, in better agreement with the experimental data than the Cartesian one.

\acknowledgments
The authors acknowledge support from Consejo Nacional de Investigaciones Cient\'{\i}ficas y T\'ecnicas de la Rep\'ublica Argentina (CONICET, Argentina) with grant PIP 844/2011 and Agencia Nacional de Promoci\'on de Cient\'{\i}fica y Tecnol\'ogica (ANPCyT, Argentina) with grant PICT 931/2012.

\end{document}